\documentclass{aa}  
\usepackage{graphicx}
\usepackage{txfonts}
\newcommand{\rhos}{\rho_\odot}
\newcommand{\rhosm}{\langle\rhos\rangle}
\newcommand{\rhop}{\rho_\oplus}
\newcommand{\rhopm}{\langle\rhop\rangle}
\newcommand{\ld}{$\:\lambda/D\:$}
\newcommand{\Strehl}{\mathcal{S}}
\newcommand{\degree}{$^{\circ}$} 

\usepackage[colorlinks=true, allcolors=blue]{hyperref}

\begin{document}

   \title{The Phase Induced Amplitude Apodizer and Nuller}

   \subtitle{High transmission, high dispersion coronagraphy at 2$\lambda/D$}

   \author{     N. Blind        \inst{1}
          \and  N. Restori      \inst{1}
          \and  B. Chazelas    \inst{1}
          \and  C. Lovis        \inst{1}
          \and  J. K\"uhn       \inst{2}
          \and C. Mordasini     \inst{3}
          }

   \institute{
        Observatoire de Gen\`eve, D\'epartement d'Astronomie, Universit\'e de Gen\`eve, Chemin Pegasi 51, 1290 Versoix, Switzerland
        \and Division of Space and Planetary Sciences, University of Bern, Sidlerstrasse 5, 3012 Bern, Switzerland
         \and Division of Space Research and Planetary Sciences, Physics Institute, University of Bern, Gesellschaftsstrasse 6, CH-3012 Bern, Switzerland
         }

   \date{Received xxx; accepted xxx}

  \abstract
  {Proxima Cen b is the prime target for the search of life around a nearby exoplanet by characterizing its atmosphere in reflected light. Due to the very high star/companion contrast ($\leq 10^{-6}$), High Dispersion Coronagraphy is the most promising technique to perform such a characterization.}
   {With a maximum separation of 37\:mas, Proxima b can be observed with a VLT in the visible. It requires a coronagraph providing high contrast ($\le 10^{-4}$) very close from the star ($\le\:$ 2\ld) over a broad spectral range ($\sim30\%$), with a high transmission of the companion ($\ge 50\%$). We look for an optimal solution that takes benefit of the properties of single-mode fibers.}
   {We introduce the Phase Induced Amplitude Apodizer and Nuller (PIAAN), a coronagraphic integral field unit, designed to feed a diffraction limited spectrograph. It uses a pupil remapping optics with moderate apodization, combined to a single mode fiber integral field unit. It exploits the properties of single mode fibers to null the star light without reducing the companion coupling. The study focuses on a proper tolerance analysis and proposes a wavefront optimization strategy. A prototype is built to demonstrate its performance.}
   {We show that the PIAAN can theoretically provide contrasts of $7\cdot10^{-7}$ and a transmission of $72\%$ at 2\ld over a bandwidth of $30\%$. A prototype is built and characterized and the proposed wavefront control strategy is also demonstrated in the lab. We reach contrast levels of $3\cdot10^{-5}$ over the full bandwidth, as expected from the tolerance analysis.}
   {We demonstrated the potential of the new PIAAN coronagraph, from simulations to prototype. Its performance will eventually be limited by the XAO capabilities. It is the main coronagraph candidate for the RISTRETTO instrument to observe Proxima Cen b on the VLT and its first technology milestone.}

   \keywords{Coronagraphy                
               }
    \maketitle

\section{INTRODUCTION}
\label{sec:intro}

We are at a pivotal moment in exoplanetary research, with the discovery of numerous Earth-like exoplanets. However, the detection and characterization of such a companion faces significant challenges due to their close angular proximity to host stars and the brightness contrast between companions and stars.

High-contrast imaging techniques, supported by eXtreme Adaptive Optics (XAO) systems, have enabled advancements in this field in the last decade. Instruments like SPHERE \citep{beuzit_2019a}, GPI \citep{macintosh_2014a}, SCExAO \citep{jovanovic_2015a}, or MAGAO-X \citep{males_2024a} use coronagraphs to suppress stellar light, allowing the faint signals of companions to be isolated. Combining high contrast with high-resolution spectroscopy, also known as High Dispersion Coronagraphy (HDC), further improves contrast by exploiting differences in spectral lines caused by factors such as Doppler velocity shifts and molecular compositions \citep{sparks_2002a, snellen_2015a}. It has already been successful in characterizing giant exoplanet atmospheres and is a key focus for projects like VLT/HiRISE \citep{vigan_2022a} and Keck/KPIC \citep{mawet_2016a}. Those coronagraphs are generally limited to a discovery zone beyond 3-4\ld\:from the star, not fully exploiting the resolving power of the largest telescopes available or to come.

One of the current golden target in exoplanet research is Proxima Cen b, an Earth like companion orbiting in the habitable zone of the closest star to us. With a maximum apparent separation of 37\:mas (i.e. 2\ld at $\lambda$=730\:nm) and a total contrast of $10^{-7}$, it is potentially observable with current 8m and 10m telescopes using HDC technique \citep{lovis_2016a, blind_2024a}. Obtaining such contrasts so close from the star require the use of an XAO delivering Strehl ratio $\ge 70\%$ in the visible, and a coronagraph providing contrast $\le 10^{-4}$ at a very small Inner Working Angle (IWA) $\le$\:2\ld in the visible. Spectral separation of the star and companion in the spectrograph brings an additional $10^{-3}$ contrast 
 \citep{snellen_2015a}. Aforementioned instruments were not designed for this restrained and difficult parameter space, so a new instrument, RISTRETTO \citep{lovis_2024a}, is being designed. Planet photons being scarce, a coronagraph providing high transmission and high contrast with low IWA is mandatory. Such coronagraphs are notoriously difficult to design and operate due to tight tolerances to low-order aberrations, tip-tilt jitter in particular that quickly degrade contrast. As noted by \citet{mawet_2012a} and \citet{por_2018a}, pupil plane only coronagraphs have unacceptably low transmission for IWA of 1-2\ld. Since the XAO provide diffraction limited images, the use of single-mode fibers is justified to feed the spectrograph. By exploiting the spatial filtering properties of these fibers, it is possible to cancel on-axis stellar light with a limited (if any) impact on the coronagraph transmission. A couple of such solutions, so called {\it nullers}, have been proposed already (Tab.~\ref{tab:cifu_list}). Among them only the Vortex Fiber Nuller (VFN) has been tested on-sky so far \citep{echeverri_2019a}, and was strongly limited by the XAO performance, although its IWA is smaller than we aim in this study.

A complete comparison of the solutions in Tab.~\ref{tab:cifu_list} is beyond the scope of this paper. We can nevertheless note that the VFN offers an annular field of view, while the SCAR and PIAAN require 6 fibers and 2 exposures to cover the same area, due to the limited field of view of single mode fibers (see Sect.~\ref{sec:piaan}). If the companion position is unknown, the VFN gets a transmission and detector readout advantage. If, on the other hand, the companion position is well known, this advantage fades, and the VFN could also suffer from additional residual speckles or background photon noise in its larger field of view.

\begin{table*}
    \centering
    \vspace{0.3cm}
    \begin{tabular}{l|crcrrrcc}
    \hline
          Coronagraph & $N_{fibers}$ & $\Delta\lambda$ & Working Angle & \multicolumn{3}{c}{Contrast}& Throughput & Jitter sensitivity\\
                      &             &                   &  [\ld] & Theory & Lab. & Sky   &          & [\ld] \\
          \hline
        VFN (l=1) \textsuperscript{1}    & 1 & $\ge$30\% & 0.9 & $\infty$ & $6\cdot10^{-5}$ & $3\cdot10^{-2}$ & 19\% &  0.01 \\
        VFN (l=2) \textsuperscript{1}    & 1 & $\ge$30\% & 1.3 & $\infty$ & $4\cdot10^{-4}$ & - & 10\% &  0.1 \\
        PLN \textsuperscript{2}    & 1 & $\ge 10\%$ & 1 & $10^{-3}$ & $10^{-2}$ & - & 60\%& n.a. \\
        SCAR360 \textsuperscript{3} & 6 & 20\% & 2 & $3\cdot10^{-5}$ & $2\cdot10^{-4}$ & - & 23\% & 0.15  \\
        Ring apodizer \textsuperscript{4} & 6 & 30\% & 2 & $2\cdot 10^{-5}$ & $\le 10^{-4}$ & - & 30\% & 0.1 \\
        PIAAN {(This work)} & 6 & 30\% & 2 & $7\cdot10^{-7}$ & $3\cdot10^{-5}$ & - &  72\% & 0.18 \\
        \hline
        \end{tabular}
        \vspace{0.3cm}
    \caption{List of coronagraphic nuller concepts. The field of view of all concepts is more or less 1\ld due to the mode size of the fibers. Contrasts are given as theoretical / measured. Throughput is only theoretical. The ring apodizer is similar to a SCAR360 that would consider a binary transmission mask. 
    Ref.:\textsuperscript{1} {\cite{ruane_2018a, echeverri_2019b, echeverri_2019a, echeverri_2023a}}; \textsuperscript{2} { \cite{xin_2022a, xin_2024a}}; \textsuperscript{3} { \cite{por_2018a, haffert_2020a}};  \textsuperscript{4} { \cite{kuhn_2022a,restori_2024a}}}
    \label{tab:cifu_list}
\end{table*}
\begin{figure}
    \includegraphics[width=\columnwidth]{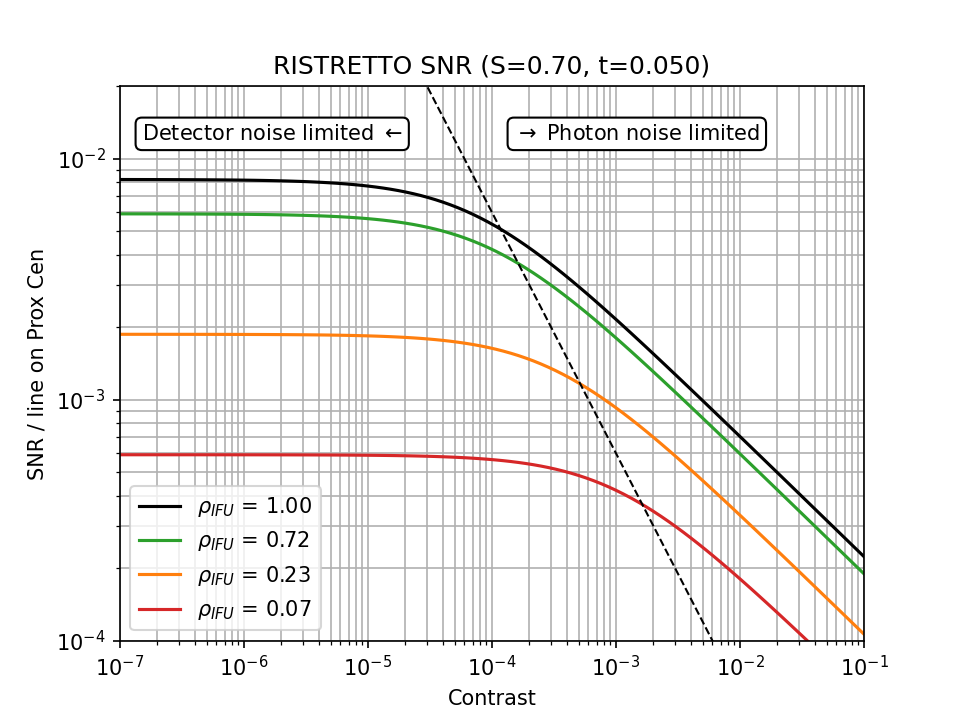}
    \caption{Expected SNR for a Proxima Cen b observation on a VLT as a function of contrast for different coronagraph transmissions at 2$\lambda/D$. We assume a total optical transmission of 5\% (excluding the coronagraph transmission), a Strehl ratio of 70\%, and $\sigma_{det}=6e^-$ for a reduced pixel in 1h of exposure time \citep{lovis_2016a}.}
    \label{fig:snr_prox_cen}
\end{figure}

The Signal-to-Noise Ratio per absorption line of the HDC technique writes as \citep{snellen_2015a}:
\begin{equation}
    SNR = \dfrac{T \rhop F_\oplus}{\sqrt{T \rhos F_\odot + \sigma^2_{det}}}
    \label{eq:snr}
\end{equation}
$T$ is the total instrument transmission. $F_\odot$  and $F_\oplus$ are the flux from the host star and companion respectively. $\rhos$ and $\rhop$ are the slit efficiency for the host star and the companion respectively. In the case of this paper, this corresponds to the coupling into a single-mode fiber feeding the spectrograph. Finally, $\sigma_{det}$ is the noise from the detector (readout \& dark). Considering observations limited by the stellar photon-noise, the SNR of HDC technique is proportional to:
\begin{equation}
    \mathrm{SNR} \varpropto \sqrt{T} \dfrac{\rhop}{\sqrt{\rhos}} \varpropto \sqrt{T} \dfrac{\sqrt{\rhop}}{\sqrt{C}},
    \label{eq:snr_shot_noise}
\end{equation}
where $C = \rhos/\rhop$ is the coronagraph contrast, at the slit or fiber level. Having a coronagraph with a 3 times higher transmission through $\rhop$ requires 10 times shorter integration times. This gain also allows to reach equal SNR for contrast requirements 3-10 times lower in this regime. Higher transmission is therefore a fundamental advantage, relaxing requirements over subsystems.

Fig.~\ref{fig:snr_prox_cen} presents the expected SNR from Eq.~\ref{eq:snr} in the case of Prox Cen b on an 8-m telescope for different contrast values $C$, assuming XAO providing $\Strehl=0.7$ (which represents a transmission term in the case of single mode fibers). It emphasizes the fundamental importance of the coronagraph transmission: an horizontal reading of the graph shows that the ultimate performance we can expect from a coronagraph with $\rho=0.23$ will be as good as a coronagraph with 3 times higher transmission ($\rho=0.72$) with a 10 times lower contrast, around $10^{-3}$. For a ground-based instrument, higher transmission also presents a fundamental benefit on the XAO design, by relaxing Strehl and low order performance requirements and by pushing magnitude limit. 

This paper presents the Phase Induced Amplitude Apodizer and Nuller (PIAAN), from the concept to demonstration. Sect.~\ref{sec:piaan} will present the concept and the theoretical solution, including an extensive tolerance analysis. Sect.~\ref{sec:wfc} will present a wavefront control strategy, mandatory to face real world operations, which is not limited to the PIAAN. In Sect.~\ref{sec:prototyping}, we present our prototyping activities, our test set-up, the characterization of the different parts, and finally demonstration of the PIAAN performance.

\section{Phase Induced Amplitude Apodizer and Nuller}
\label{sec:piaan}

\subsection{Concept}

The idea of the Phase Induced Amplitude Apodizer and Nuller (PIAAN) is derived from speckle nulling techniques that exploit the spatial filtering properties of Single Mode Fibers (SMF) to cancel stellar light coupled into a fiber with a minimal transmission cost \citep{mennesson_2006a}. In such fiber, the coupling efficiency $\rho$ depends on the overlap integral between the SMF fundamental mode $E_{f}$ and the electric field at focus $E_o$, which is easily computed in the SMF far-field \citep{marcuse_1973a}:
\begin{equation}
    \rho = \dfrac{\left| \int E_o E_{f} dS \right|^2}{\int |E_o|^2 dS \int |E_{f}|^2 dS}  
\end{equation}
$E_f$ being unique with a quasi-gaussian shape and the core of an Airy PSF being a good match, $\rho$ is maximized when the PSF is centered onto the fiber. Otherwise, coupling is reduced, with some peculiar positions presenting a perfect null, i.e. $\rho = 0$. This generally happens when the fiber is between two diffraction rings or speckles, whose phase is opposite over the fiber mode, thus canceling each other in the overlap integral.

Coronagraphs like the SCAR \citep{por_2018a} or the VFN \citep{ruane_2018a} rely on this property to provide low stellar coupling $\rhos$ very close from a star, between 0.5 and 3\ld. This however comes with limited companion coupling $\rhop$ and/or small bandwidth. 

\begin{figure}
    \centering
    \includegraphics[width=0.5\textwidth]{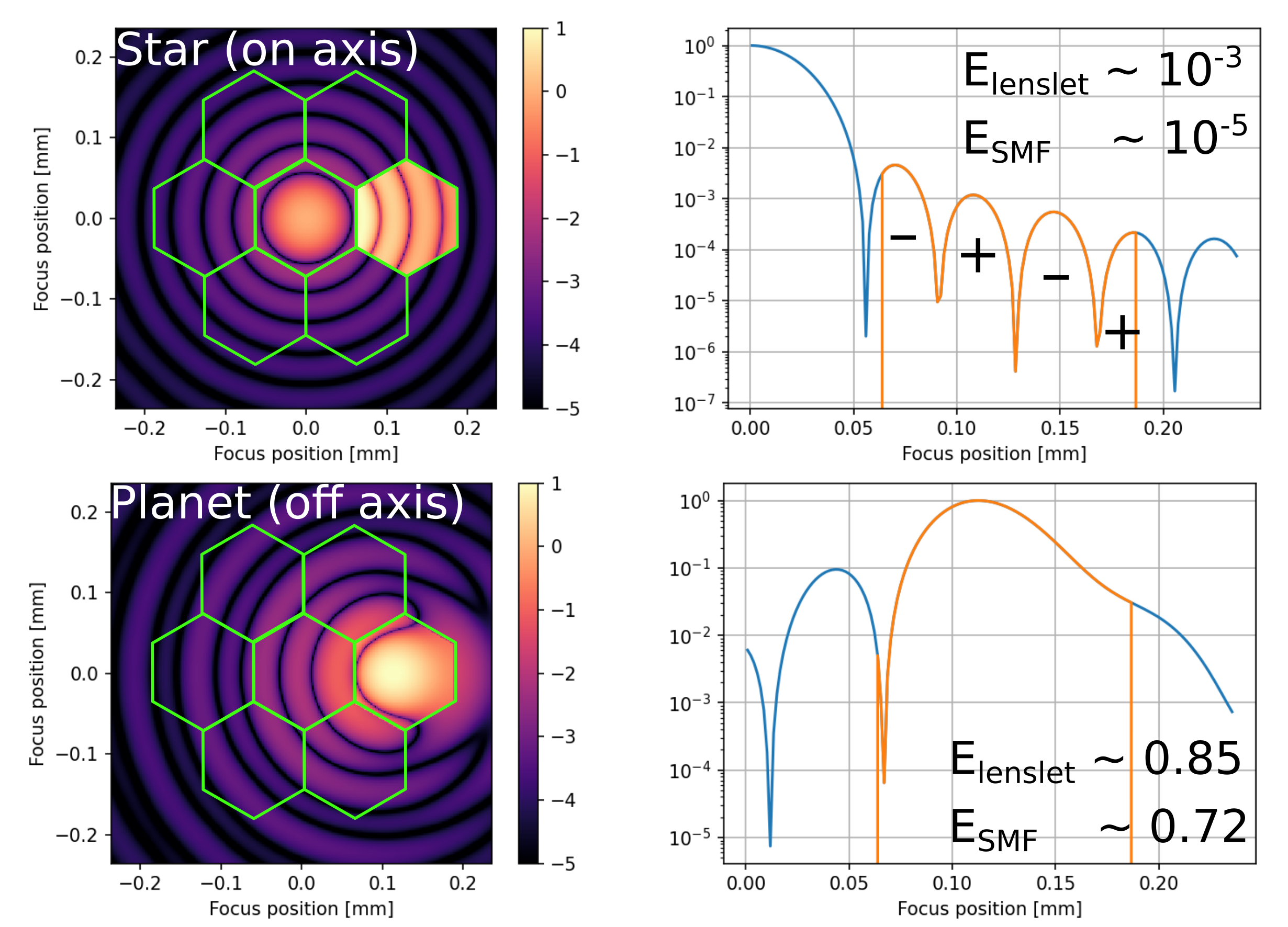}
    \caption{Working principle of the PIAA Nuller. Top: coupling of the star (on-axis). Bottom: coupling of the companion at 2\ld. Left: PIAA PSF and overlayed IFU at 620nm, with the 'companion' lenslet showcased. Right: Horizontal cut view of the PSF in log scale, with yellow part showing the electric field coupled to the fiber at the companion position.}
    \label{fig:piaan_concept}
\end{figure}

The PIAAN is a coronagraphic integral field unit similar to the SCAR in the working principle, providing both high contrast and high transmission at a separation of 1.5 to 2.5\ld. It uses two main components: 
\begin{itemize}
    \item A PIAA optics that only partially apodizes the pupil, generating a PSF with rings of lower amplitude and higher frequency. Compared to a standard VLT PSF, the rings are 10-40 times fainter, and 2 to 3 times more numerous.
    \item An Integral Field Unit made of 7 hexagonal lenslets feeding 7 SMFs.
\end{itemize}
Because phase is changing sign at every consecutive diffraction ring in the focal plane, they cancel each other efficiently over the fiber mode, allowing nulls $\rho \sim 10^{-5} - 10^{-6}$ (Fig.~\ref{fig:piaan_concept}, top). This also enhances the bandwidth and improves the robustness to wavefront error, tip-tilt jitter in particular. 

Another property of the PIAA optics is to increase the match between the SMF mode and PSF core. While for a VLT, $\rho \le 70\%$ in absence of aberrations \citep{ruilier_1998a}, the PIAA potentially allows to reach $\rho = 100\%$ by removing the diffraction rings and concentrating light into a gaussian core \citep{jovanovic_2017a}. This benefit is however valid only if the fiber is placed on the PIAA optical axis. Due to the non-linear phase transformation, off-axis objects quickly get aberrated with a coma-like distortion (corresponding to the transformation of the tip-tilt phase by the PIAA), finally degrading $\rhop$. The degradation can be very fast, since 1\ld tip-tilt corresponds to 2$\pi$ phase. In the case of the PIAAN, the partial apodization is more gentle, the distortion not as strong leading to an off-axis Strehl degradation of about 15\%, so that the coupling is still high at 2\ld (Fig.~\ref{fig:piaan_concept}, bottom).

\begin{figure}
    \centering
    \includegraphics[width=\linewidth, trim={0cm 12.5cm 13cm 0cm}, clip]{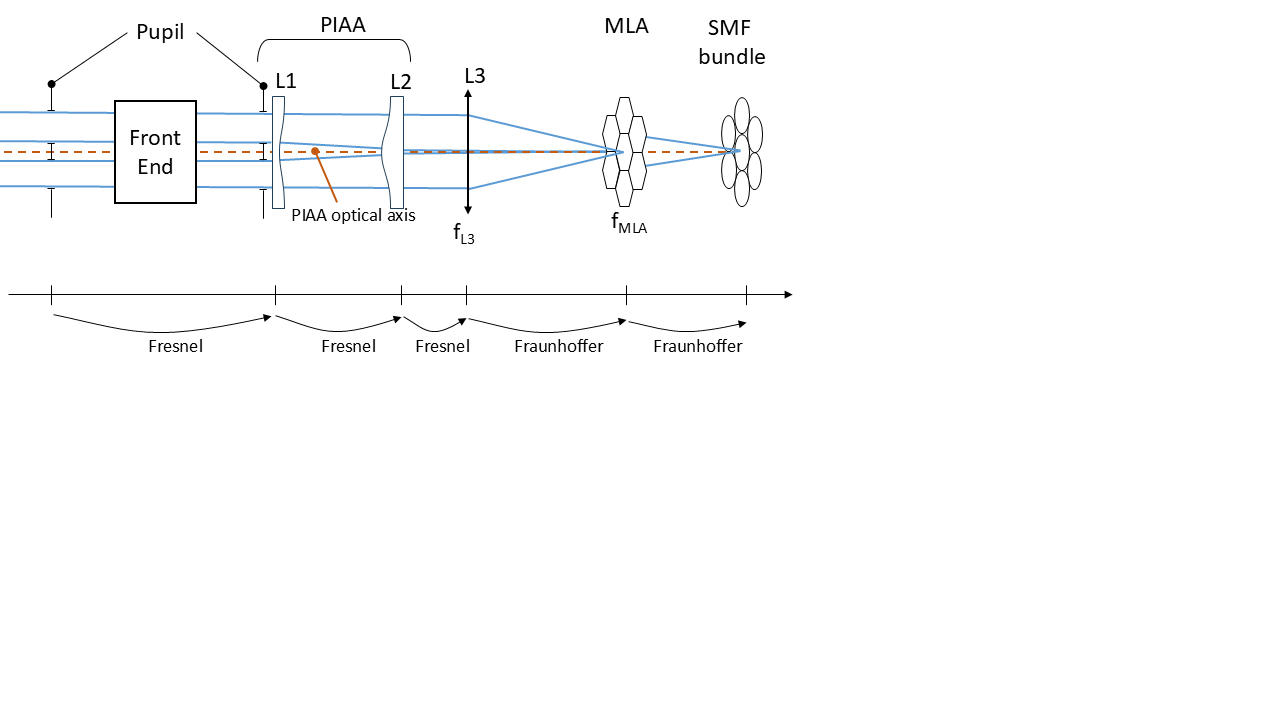}
    \caption{Schematics of the PIAAN simulation. The Front-End part is used to inject residual XAO phase screen and to simulate an optical relay with realistic optics and off-pupil aberrations. The MLA can be simulated with data from manufacturers if required.}
    \label{fig:piaa_sim_scheme}
\end{figure}

\begin{figure*}
    \centering
    \includegraphics[width=1\linewidth]{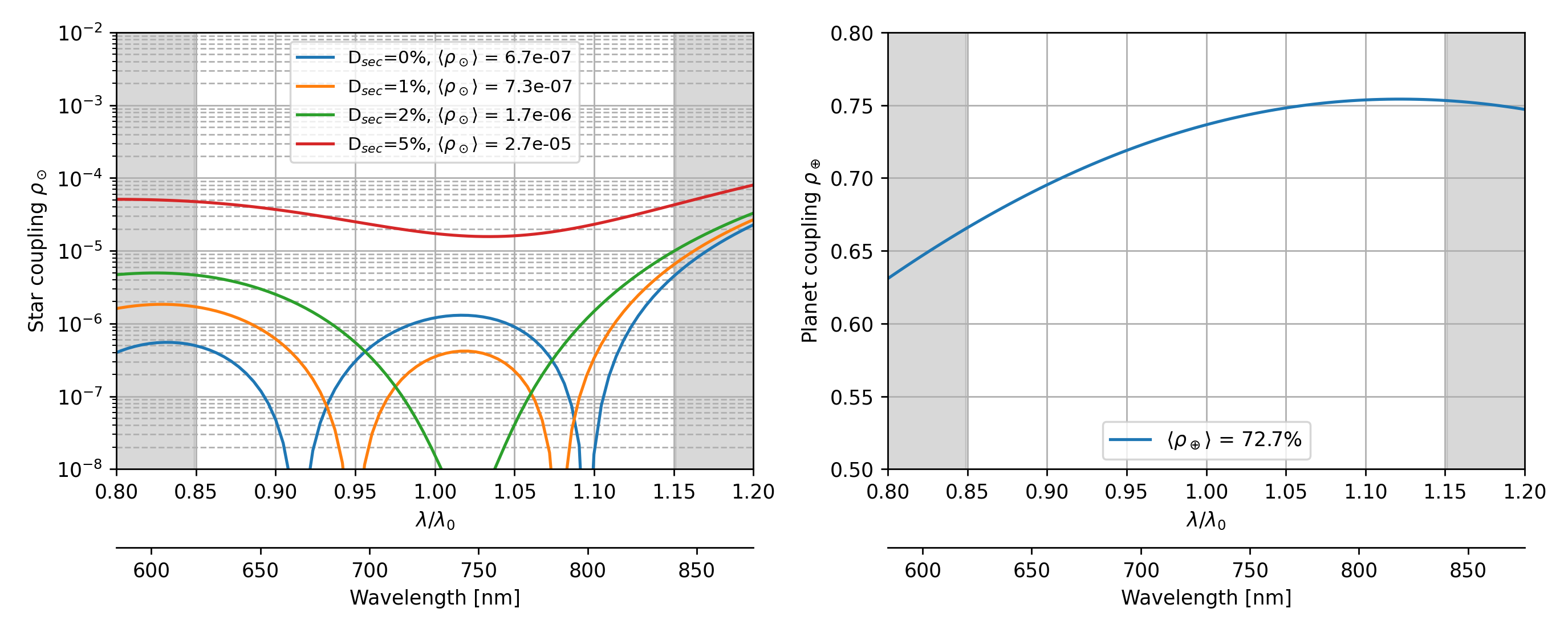}
    \vspace{-0.cm}
    \caption{Expected performance for a PIAAN optimizing detection of a companion at 2\ld with a 30\% bandwidth, i.e. in the RISTRETTO case between 620 and 840\:nm. Behavior outside the optimization band is shown in the light gray areas. Left: star null $\rhos$, showing the influence of the residual secondary obstruction. Right: companion coupling, mostly independent of the residual obstruction.}
    \label{fig:piaa_perf_geo}
\end{figure*}

\subsection{Model}

The PIAAN is simulated using the Hcipy Python library \citep{por_2018a}, from telescope to fiber coupling. The PIAA transformation and optics surfaces are computed based on ray-tracing and energy conservation principles \citep{guyon_2005a}. We first developed a purely geometrical model, where we compute and apply the desired gaussian apodization of the electric field (including a reduction in the secondary obstruction). Using entrance and exit rays coordinates, we build a function that remaps the entrance phase according to the PIAA transformation. This model is equivalent to consider an infinitely thin optics, neglecting physical propagation between surfaces. This model offers pretty fast computation speed and was therefore used for the global parametric exploration of Sect.~\ref{sec:optim_perf}. A scheme of this model is shown in Fig.~\ref{fig:piaa_sim_scheme}.

We also developed a physical model to properly design and study real optics. Sag of each optical surface is extrapolated from the geometrical ray tracing, and Fresnel propagation is used between surfaces. We considered 3 possible configurations: a set of 2 mirrors as in the original paper \citep{guyon_2005a}; a set of 2 flat-aspherical refractive optics as is more commonly done nowadays; and a novel aspherical-aspherical refractive optics (so-called rod optics). We only considered CaF$_2$ glass, whose chromatism is also taken into account in the propagation.

One of the main limitation to produce such an optics is the minimum curvature radius achievable by the diamond turning tools. In this sense, the rod design presents the drawback of requiring sag and curvature radii 6-7 times higher than the mirror one, or 1.5 times higher than a double lens design on the 1$^{st}$ surface, which present the strongest deviation. On the other hand, the design offers a proportionally lower sensitivity to sag and manufacturing errors, leading to higher optical quality. Another advantage is that the optics does not require internal alignment, which proved advantageous.
For our solutions, the physical model requires at least 600 pixels across the pupil to sample well the optics phase transformation at $\lambda$=600\:nm. Computation time for 7 fibers at a single wavelength is on the order of 1 to 2\:seconds on a standard laptop.

Apart from the PIAA, wave is generally propagated from pupil to focus with Fraunhofer propagator. Fresnel propagation is nevertheless used for tolerancing purposes or when a more accurate model can be implemented (e.g. for the microlens array), to study potential impact of Talbot effects, etc. Aberrations can be injected in the pupil (e.g. atmospheric turbulence) or outside (e.g. lenses) to evaluate performance in real conditions or to study wavefront control strategies and performance. Various defects on the IFU have also been implemented, i.e. independent near- and far-field misalignments to every lenslets.

\subsection{Optimization}
\label{sec:optim_perf}
The performance of the PIAAN depends on 4 main parameters:
\begin{enumerate}
    \item {\bf PIAA apodization} -- An aggressive PIAA design can theoretically attenuate diffraction rings to $10^{-10}$ contrast level \citep{guyon_2005a}, not requiring nulling at all. Such solutions are however not yet feasible or too complex to design. A sweet spot exists with a partial apodization, which generates a set of faint and high frequency rings that produce some null over singlemode fibers. We investigated different apodization shapes (gaussian and prolate mostly), but could not observe significant differences between best solutions, and decided to stick to the simple gaussian one.
    \item {\bf Central obstruction filling factor} -- A higher telescope central obscuration leads to stronger diffraction rings close from the PSF core. The PIAA is therefore also optimized to fill the secondary obstruction. Fully removing it leads to better performance, while keeping it higher than 0.05\:$D_{tel}$ severely limits the performance (Fig.~\ref{fig:piaa_perf_geo}, left).  
    \item {\bf Lenslet size and focal ratio} -- The lenslet constitutes a spatial filter in the focal plane and is used to optimize primarily the null level. With new 3D printing technologies, different lenslet shapes can be considered on the external ring. However, simulation with hexagonal and semi-circular lenslets led only to marginal differences, so we did not explore shape further. For the rest of the paper, we only consider hexagonal lenslets. Note also that the lenslet size or focal ratio must be adapted to the PIAA apodization, since the beam compression generates a non-linear magnification at the focal plane.
    \item {\bf Lenslet focal lens} -- It mostly optimizes coupling of the companion, but allows some fine tuning of the null level. 
\end{enumerate}
The optimization has been performed with the geometrical model via a simple grid exploration over the 4 parameters, with the largest exploration on the apodization strength and lenslet size, the main contributors to the performance. The best solutions are optimized for providing the highest SNR in the observation regime limited by the stellar photon noise, i.e. by maximizing Eq.~\ref{eq:snr_shot_noise}, over a bandwidth of 30\%. We also only consider here a VLT pupil, which does not have an impact on the results with the geometrical model.

Note that we can consider coupling the focal plane to the far-field or near-field of the SMF. Because the SMF mode field diameter is proportional to wavelength, like the diffraction limit PSF, it is generally better to perform near-field coupling to maintain high coupling over a broad range. However, when the focal plane is filtered by a lenslet, this is not true anymore. With a lenslet size of 2\ld, near- and far-field coupling options perform nearly in the same way. Since in practice, it is easier to couple to far-field, this is the only solution considered in the rest of the paper.

\begin{figure}[t]
    \centering
    \includegraphics[width=0.48\textwidth]{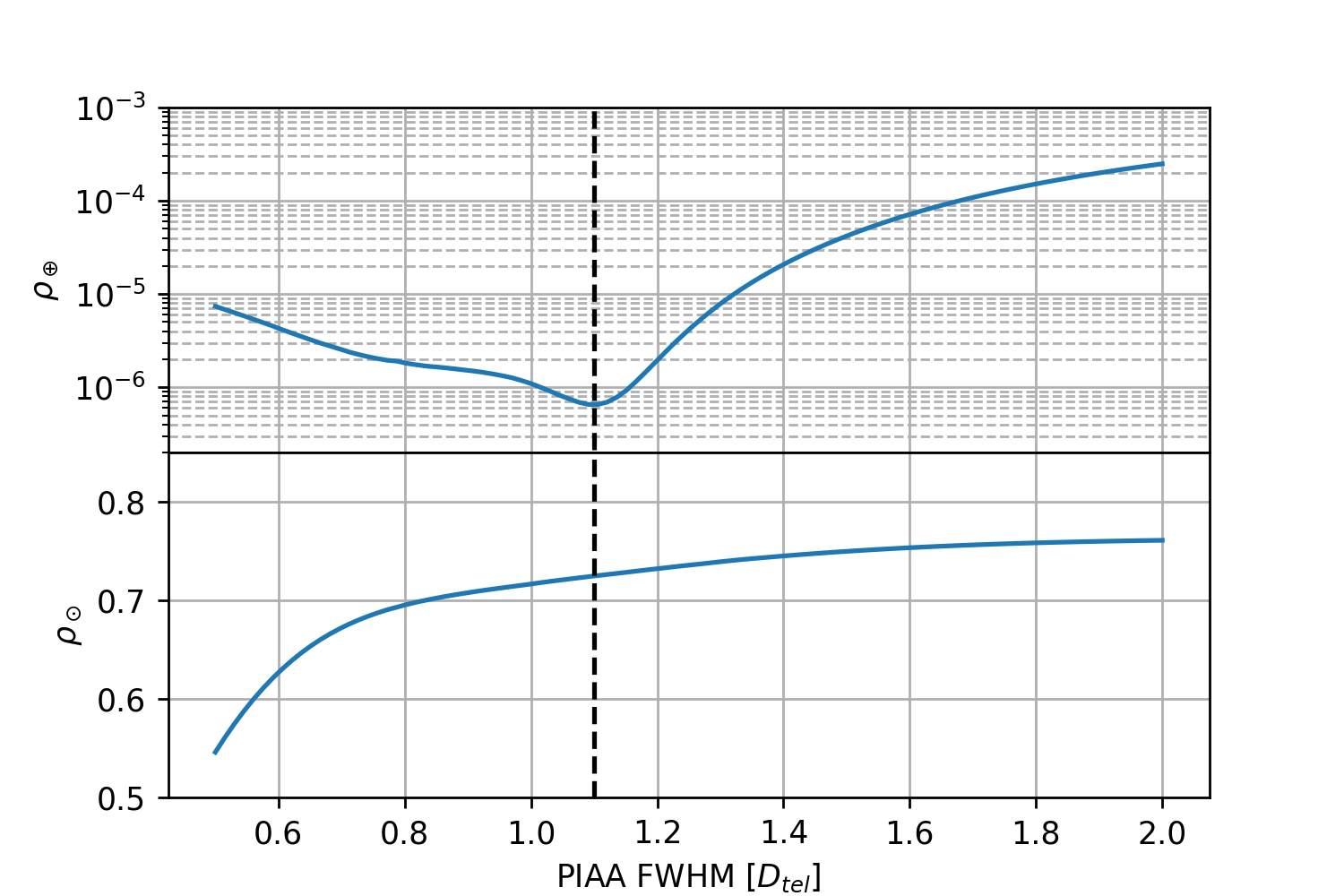}
    \caption{Influence of the PIAA apodization on the PIAAN performance. Coupling are averaged for 30\% bandwidth. Optimal solution is shown with the dashed line.}
    \label{fig:piaan_params}
\end{figure}

\begin{figure*}
    \centering
    \includegraphics[width=.95\textwidth]{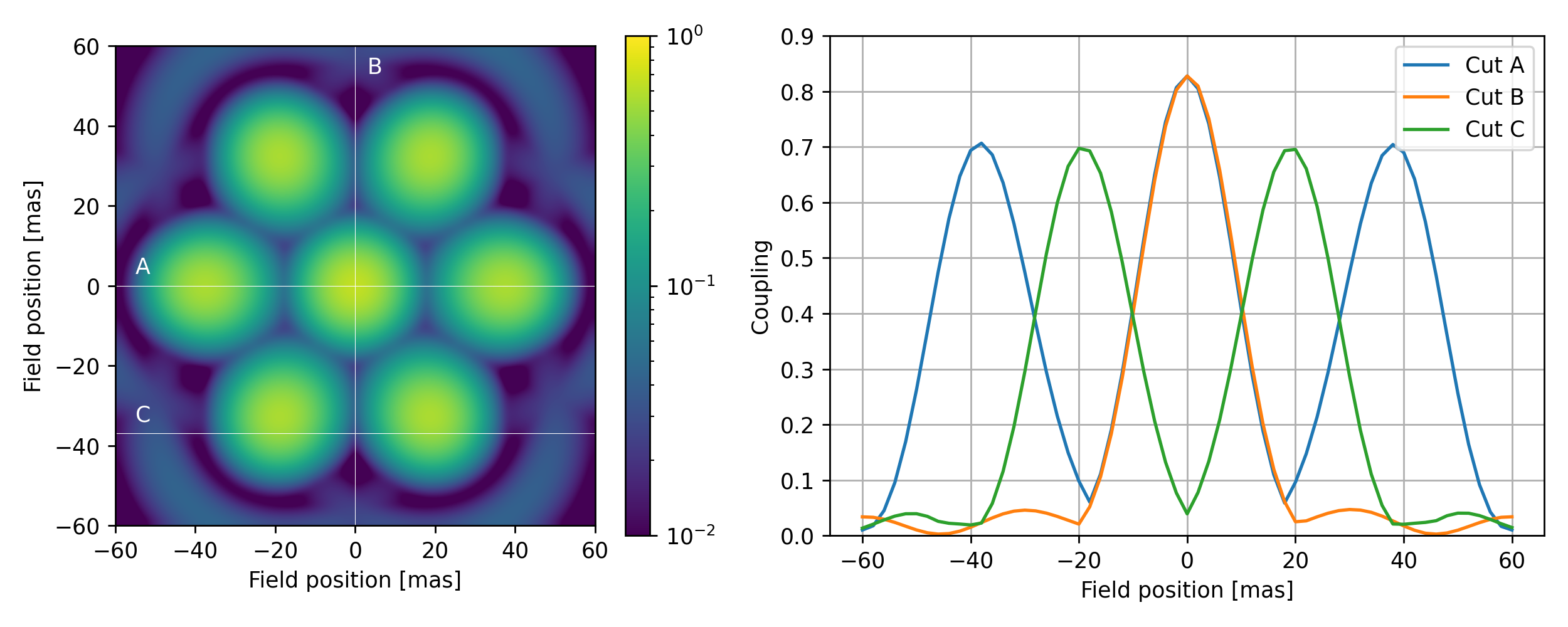}
    \caption{Left: Transmission of the PIAAN over the 7 fibers as a function of the object position. Right: 3 cuts through the left map showing the fast loss of transmission if the companion is not centered, and in particular the blind zones between fibers with transmission dropping to zero (cut B \& C).}
    \vspace{-0.2cm}
    \label{fig:piaan_throughput}
\end{figure*}

\begin{figure*}
    \centering
    \includegraphics[width=1\linewidth]{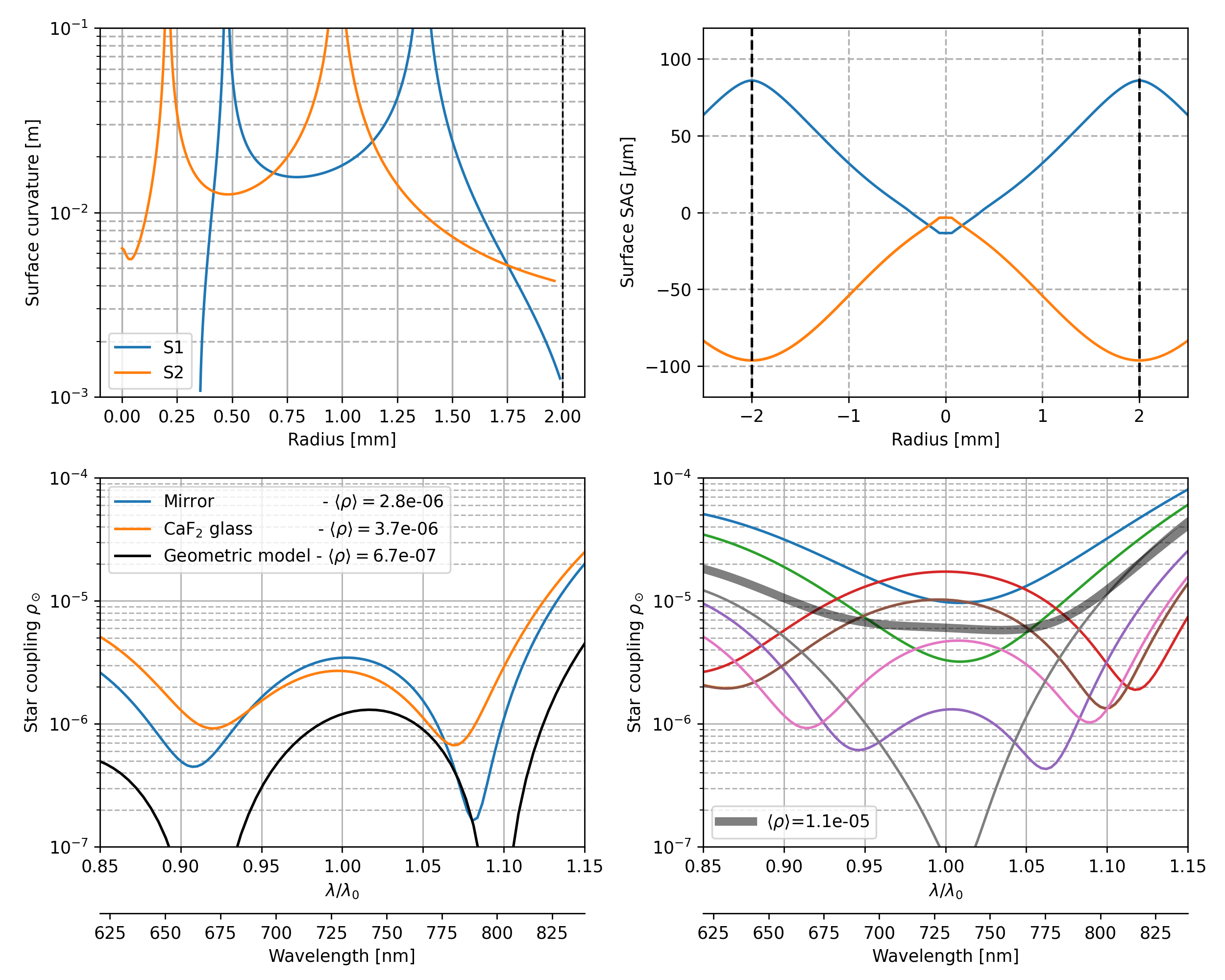}
    \vspace{-0.5cm}
    \caption{Design and performance of a real PIAAN optics. Top: curvature radius and sag of both aspheres for a CaF$_2$ rod design. Bottom, left: star coupling, comparing geometrical model to physical ones. Bottom, right: star coupling in presence of VLT spiders for different lenslet position angle relative to the pupil in color. Average over all orientations in gray.}
    \label{fig:piaa_perf_fresnel}
\end{figure*}

\begin{figure*}
    \centering
    \includegraphics[width=1.\textwidth]{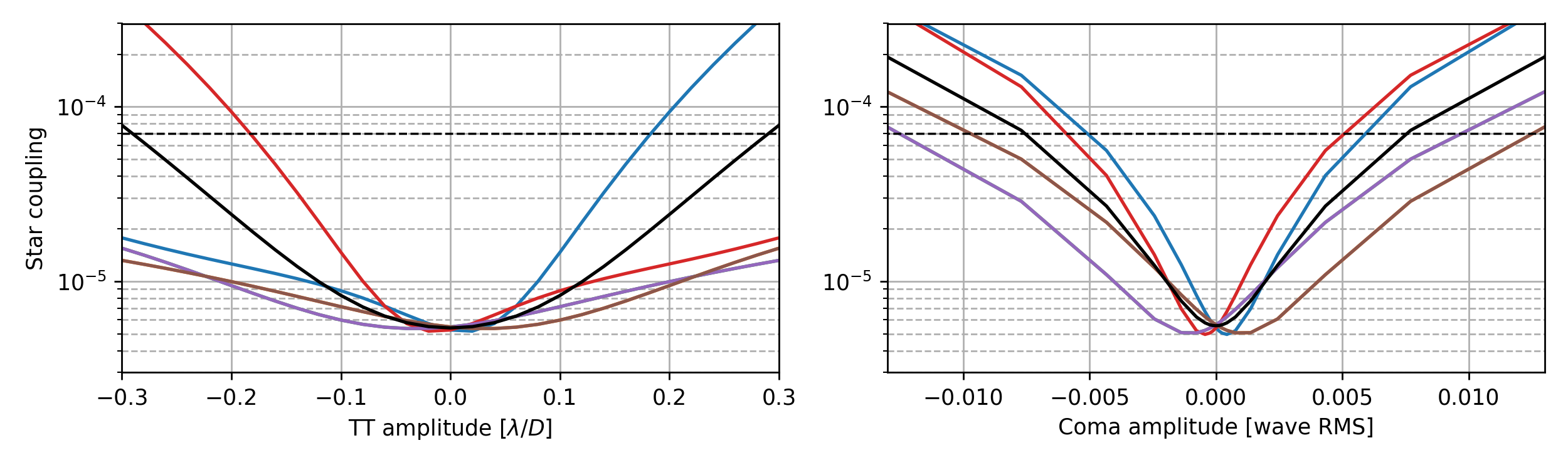}
    \caption{Sensitivity of null to tip (left) and coma (right) for 6 fibers in color, and averaged in black. Tolerance limit is shown with dashed black line.}
    \label{fig:tt_coma_sensitivity}
\end{figure*}

Fig.~\ref{fig:piaa_perf_geo} presents the performance of the best solution, providing $\rhopm = 72\%$ transmission at 2\ld, and stellar nulling $\rhos \le 10^{-5}$ at all wavelengths in the 30\% bandwidth (or even $\rhos \le 10^{-6}$ for bluer wavelength for a slight cost in transmission), and an average of $\rhosm = 7\cdot10^{-7}$. On the red side of the design band, contrast remains high but SMF cut-off wavelength is soon reached. We did not study the case of endlessly singlemode photonic fibers. On the blue side, contrast degrades due to the central lobe extending into the lenslet. We show in Fig.~\ref{fig:piaan_params} the influence of the apodization factor on the performance around the optimum solution. We observe in particular the rapid increase in off-axis aberration (plus, to a lower level, the increasing PIAA magnification), degrading $\rhop$ for FWHM $\le 0.8 D_{tel}$, while the star nulling also starts to degrade. Note that the solution at FWHM$=0.5D_{tel}$ is already very aggressive regarding manufacturing, with minimum curvature radius 10 times smaller than our optimum solution. 

It is well known that such coronagraphs have very tight tolerances, especially regarding low order wavefront error (Sect.\ref{sec:tolerances}). For all configurations, we also performed a first tolerance analysis over the first 10 Zernike modes, to check whether a configuration or parameter improved technical feasibility of the coronagraph or could even relax XAO requirements. We could only notice that solutions achieving better nulls are slightly more tolerant, but at a marginal level.

Fig.~\ref{fig:piaan_throughput} presents the transmission map of the PIAAN. The off-axis coupling is only 85\% from the central fiber due to the off-axis aberration. As expected from SMF, the transmission drops quickly if the object is not centered, with 50\% loss at $\pm$\:0.5\:\ld ($\pm$\:9\:mas for a VLT in the visible). If the companion falls between 2 spaxels, transmission drops nearly to 0 due to the mode mismatch resulting from the spatial filtering by the lenslet (Fig.~\ref{fig:piaan_throughput}, cuts B \& C). Since the position of the companion is not necessarily known, two exposures are required to homogenize the transmission, with the IFU rotated by 30\degree. This can be accounted for as a transmission loss of 50\% due to the doubled integration time. The same argument and strategy holds for a spaxel with lower performance: a spaxel with 0 transmission would account for 1/6 transmission loss with proper observing strategy. There is therefore a strong interest in determining where the companion is by other means, to place it directly on the best spaxel, and even optimize the PIAAN working point for it.

\subsection{Final design}
\label{fig:final_perf}

The geometrical solution is now injected in the physical model. Contrast performance degrades due to the Fresnel diffraction between both lenses, and therefore a real optics design requires a final adjustment step to determine adequate pupil size and aspheres separation. To ensure we get feasible optics, we considered curvature radius of aspheres above 1\:mm only. Considering a pupil size of 4\:mm, the minimal optics thickness with this constraint reaches 20\:mm for a rod design. Our best solution reaches $\rhopm = 72\%$ and $\rhosm = 3-4\cdot10^{-6}$ over the 30\% bandwidth. A mirror solution provides marginally better performance for our application. At such level, the rod design starts also to be limited by glass chromatism, which adds, to first order, a small defocus aberration apart from the design wavelength. The diffraction acts like a chromatic "background", limiting $\rhos$ to about $10^{-7}$-$10^{-6}$ around the original nulls (Fig.\ref{fig:piaa_perf_fresnel}, bottom left).

The PIAAN solution relies on a rotational symmetry of the PSF, so that diffraction from telescope spiders degrades the null performance to $\rhosm=1.1\cdot10^{-5}$ (Fig.~\ref{fig:piaa_perf_fresnel}, bottom right). We can observe the strong orientation and wavelength dependency of $\rhos$, i.e. of the SNR. PIAA optics being generally diamond turned aspheres, compensating spiders is not studied as part of the PIAA optical design. Solutions have already been proposed to fill the spider obstruction \citep{lozi_2009a}. They do not directly symmetrize the pupil illumination in case of spiders with angles different than 90\degree, but a mask could be used to redefine the pupil for a small transmission and spatial resolution cost. We also believe that a small phase offset could symmetrize the nulls, with an optimization similar to the SCAR concept, but considering a very small correction since diffraction rings are a lot fainter. Considering a ground-based application, the null degradation due to spiders is in practice limited due to low order XAO residuals (see Sect.\ref{sec:wfe_tol}).

\subsection{Tolerances}
\label{sec:tolerances}

We now analyze tolerances with respect to manufacturing, alignment or wavefront errors. Our specification is set for the case of an observation of Proxima b with RISTRETTO, i.e. $\rhop \ge 50\%$ and $\rhos \le 10^{-4}$ according to \citet{lovis_2016a}. Tolerance based on SNR could also be used, but results vary at the margin, and are dominated by the degradation of $\rhos$: so we focus on the latter term for this analysis. Finally, to reach this performance on sky, the error budget is split in two equal terms: a static one (manufacturing and alignment) and a dynamic one (XAO residual). This let us with the following specifications:

\begin{center}
$\rhos^{\mathrm{static}} \le 7\cdot10^{-5}   \qquad\qquad\qquad 
    \rhos^{\mathrm{dynamic}} \le 7\cdot10^{-5} $
\end{center}

\begin{figure*}
    \centering
    \includegraphics[width=0.45\linewidth]{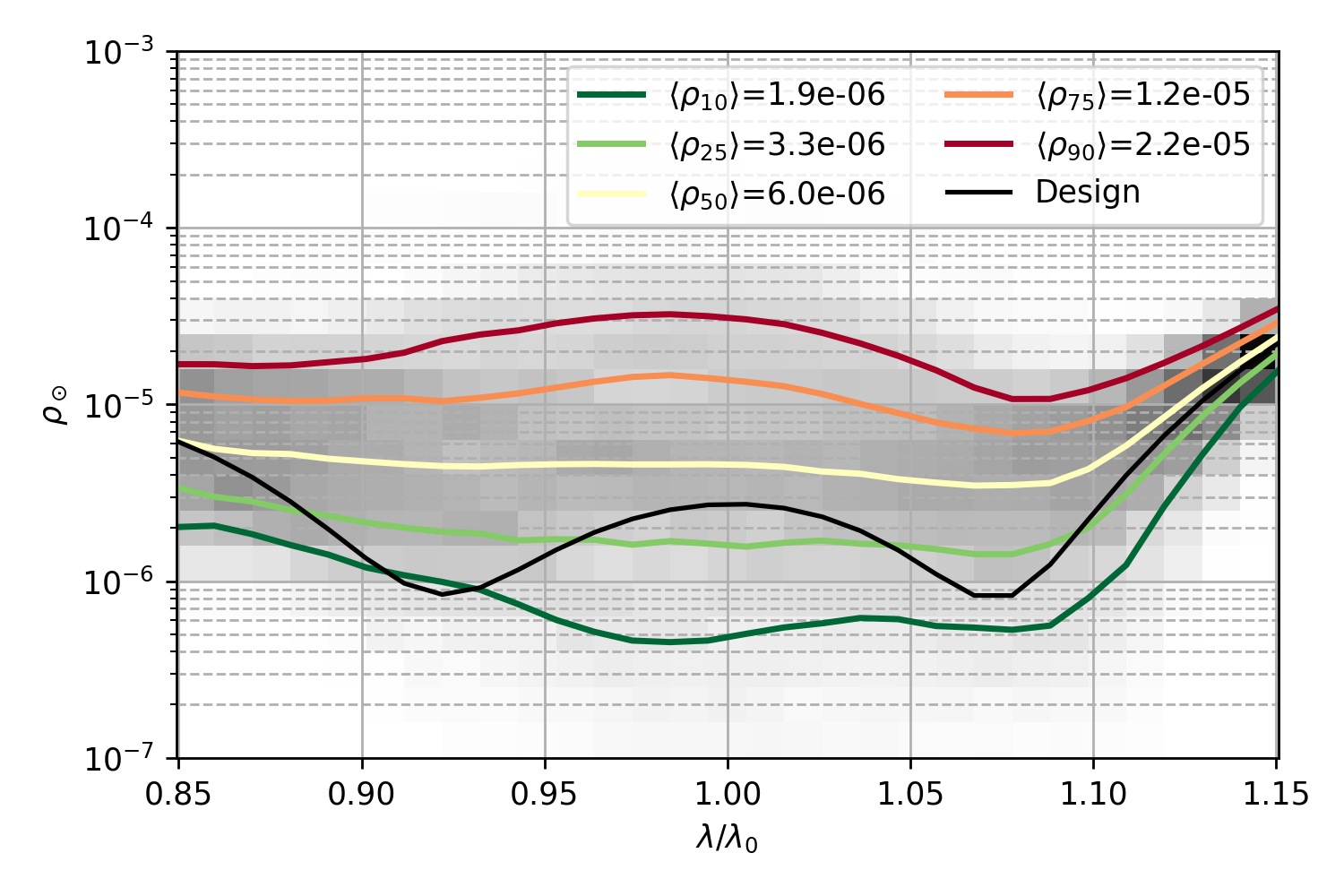}
    \includegraphics[width=0.45\linewidth]{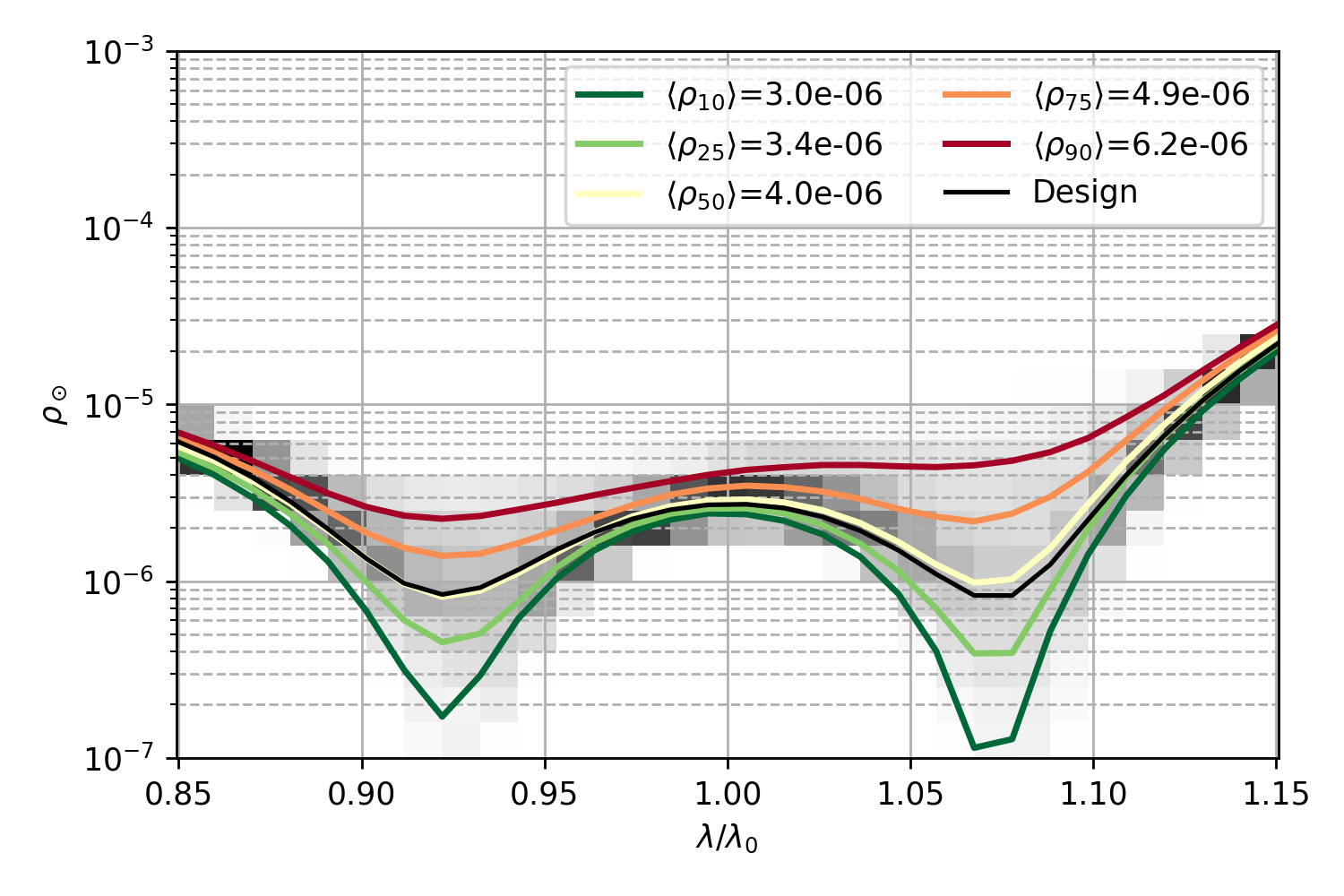}
    \caption{Probability distribution of the PIAAN performance considering manufacturing tolerances. There is not optical aberration in this system, not even on PIAA optics. Gray area is the probability density function for each wavelength. Curves then represent the 10, 25, 50, 75 and 90-percentile of the realizations for each wavelength. They do not represent a likelihood to obtain an actual contrast curve. Left: all tolerances included as presented in Tab.~\ref{tab:manufacturing_errors}. Right: a perfect IFU is considered. 
    }
\label{fig:mcmc-tolerance}
\end{figure*}

\subsubsection{Tip-tilt sensitivity}
\label{sec:tt_tol}

One of the main disturbance in a telescope is due to residual tip-tilt, either from atmosphere residuals and chromatism, or from vibrations. For the fiber in the direction of the tip, we observe a tolerance of $\pm\:0.18$\ld, i.e. $\pm\:3.5$\:mas on a VLT at 750\:nm  (Fig.~\ref{fig:tt_coma_sensitivity}, left). The degradation happens when the PSF core starts to leak into the lenslet and is not well balanced by the rings. $\rhos$ is still maintained $\le 7\cdot 10^{-5}$ in average for all wavelengths and all fibers up to $\pm$\:0.28\ld, for which companion coupling drops from 72\% to 60\%. Such large tolerance will strongly help in operations and design, by limiting the impact of tip-tilt jitter and relaxing ADC design.
 
That also means the PIAAN is relatively insensitive to the star apparent diameter. A relatively large star like Proxima Cen, with a diameter of $\sim$\:1\:mas, will not degrade the nuller performance on a VLT or even an ELT.

\subsubsection{Coma tolerance}
\label{sec:coma_tol}

Coma is an aberration easily created by optical misalignment. It also appears as the most sensitive aberration among low order Zernike modes. We actually observe an extremely tight tolerance range of about $\pm\:\lambda/200$ RMS averaged over the science bandwidth for spaxels in the direction of the aberration (Fig.~\ref{fig:tt_coma_sensitivity}, right). Note that the same analysis performed on a perfect coronagraph leads to the exact same curves and tolerance, demonstrating this is not solely a property of the PIAAN but of the requirements at such small working angle.


%
\begin{table*}[]
    \centering
    \begin{tabular}{l | ccl}
         \hline
         \hline
         Parameters & Tolerance & MCMC value & Comment \\
         \hline 
         \multicolumn{4}{l}{{\bf Wavefront errors}}\\
         \hline
         Tip-tilt   & $\pm$ 0.18\:\ld  &     -        & Jitter, ADC residual \\
         Coma       & $\lambda/200$ RMS &    -         & \\
         XAO residual   & $\lambda/75$ RMS in 3 cycles  & - &\\
         \hline 
         \multicolumn{3}{l}{{\bf PIAA optics}}\\
         \hline
         L1-L2 distance     & $\pm$ 0.30\% & - & Can be measured before diamond turning\\
         L1-L2 misalignment & $\pm$ 0.010\:$D_{pupil}$ & $\pm$\:0.01\:$D_{pupil}$\:PTP & Limited by chromatism in rod design \\         
         Pupil misalignment & $\pm$ 0.025\:$D_{pupil}$  &  $\pm$\:0.005\:$D_{pupil}$\:PTP & Easy to align from PIAA pupil image\\
         Pupil magnification & $\pm$ 0.020\:$D_{pupil}$ & - & Pupil can be redefined at PIAA level\\
         Pupil mask to PIAA distance & $\pm$3\:cm & $\pm$1\:cm PTP & \\
         \hline 
         \multicolumn{3}{l}{{\bf IFU}}\\
         \hline
         f$_{L3}$ \& MLA size & $\pm$ 5\% & 1\% RMS & As measured \\
         L3 defocus & $\lambda/20$ RMS & $\lambda/20$ RMS & \\
         Alignment to PIAA axis & $\pm$ 0.3\:$\lambda$/D& - & \\
         f$_{MLA}$ \& SMF MFD & $\pm$ 10\% & 3\% RMS & As measured in Sect.~\ref{sec:smf630hp} \\
         Near-field alignment & $\pm$ 0.1 MFD & 0.03 MFD RMS & \\  
         Far-field alignment &  $\pm$ 1\degree & 0.06\degree\:RMS & As measured in Sect.~\ref{sec:bundle} \\     
         \hline
    \end{tabular}
    \vspace{0.3cm}
    \caption{Summary of PIAAN tolerances. Tolerance value is given as the limit ensuring $\rhosm \le 7\cdot10^{-5}$ for a single defect. We also give parameters of MCMC analysis with PTP value for a uniform distribution, and RMS for a gaussian one. Those set to 0 are believed to be under control.}
    \label{tab:manufacturing_errors}
\end{table*}

\begin{figure*}
    \centering
    \includegraphics[height=6.5cm, keepaspectratio, trim={0.3cm 0 0.3cm 0}, clip]{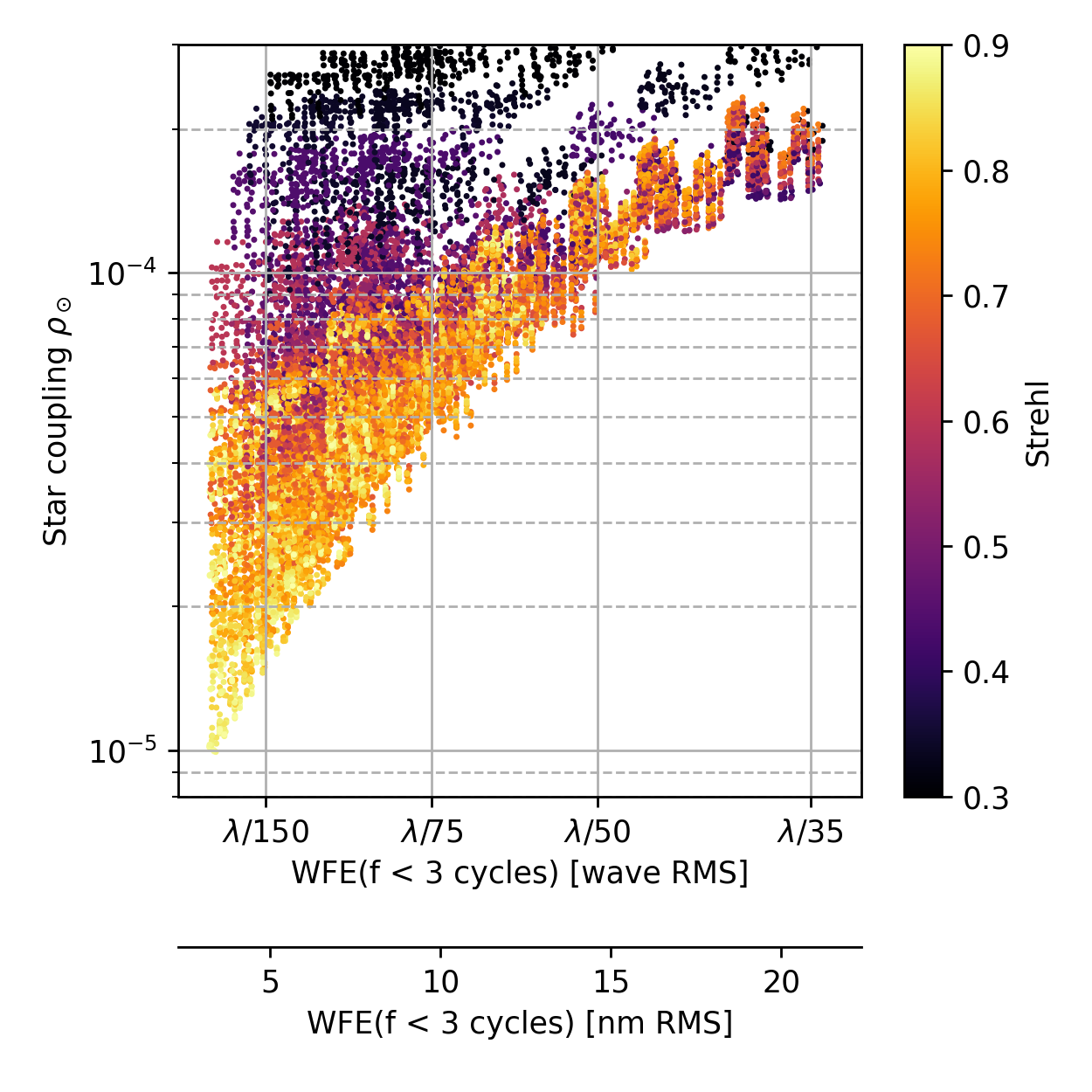}
    \includegraphics[height=6.5cm, keepaspectratio, trim={0cm 0 0.3cm 0}, clip]{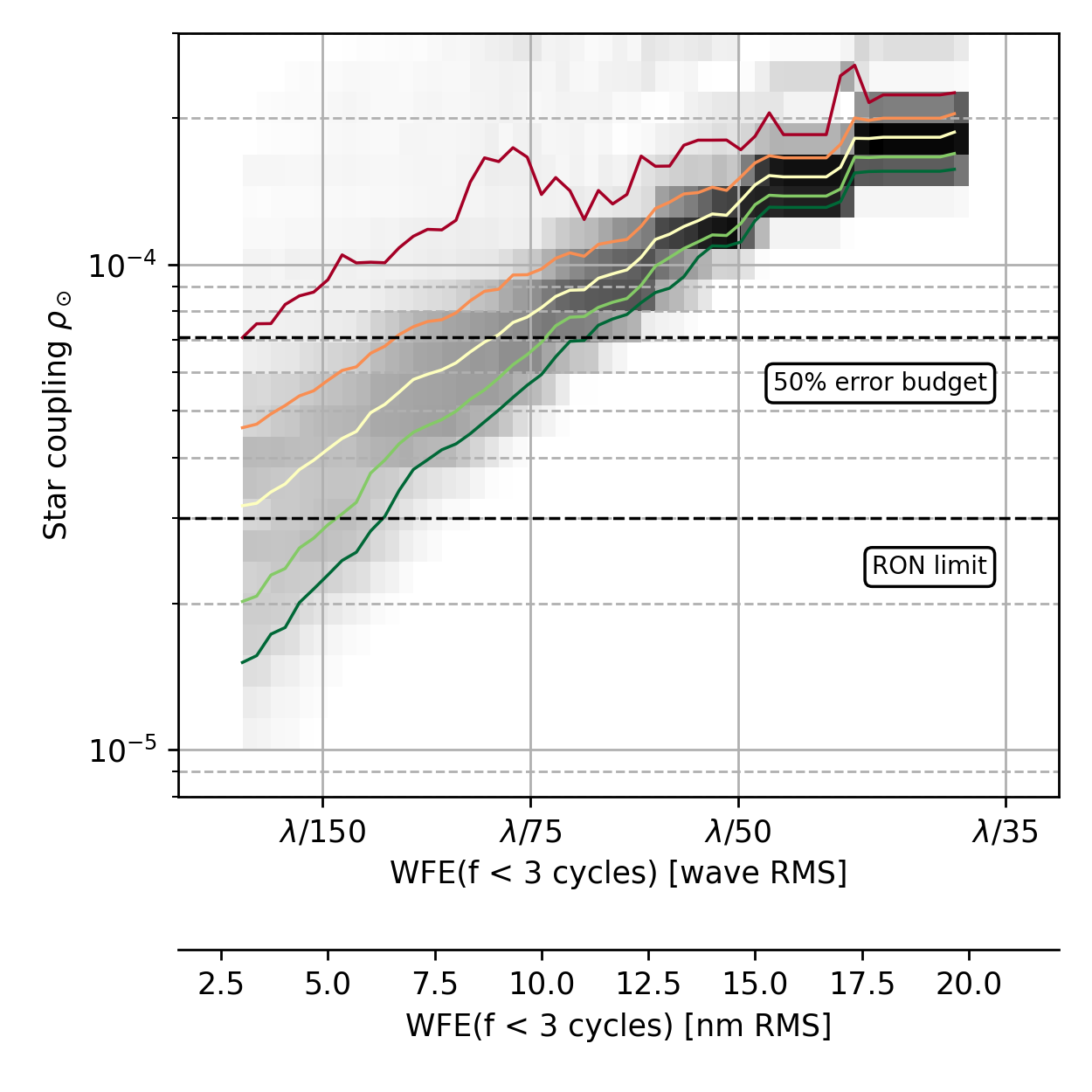}
    \includegraphics[height=6.5cm, keepaspectratio, trim={2cm 0 0.3cm 0}, clip]{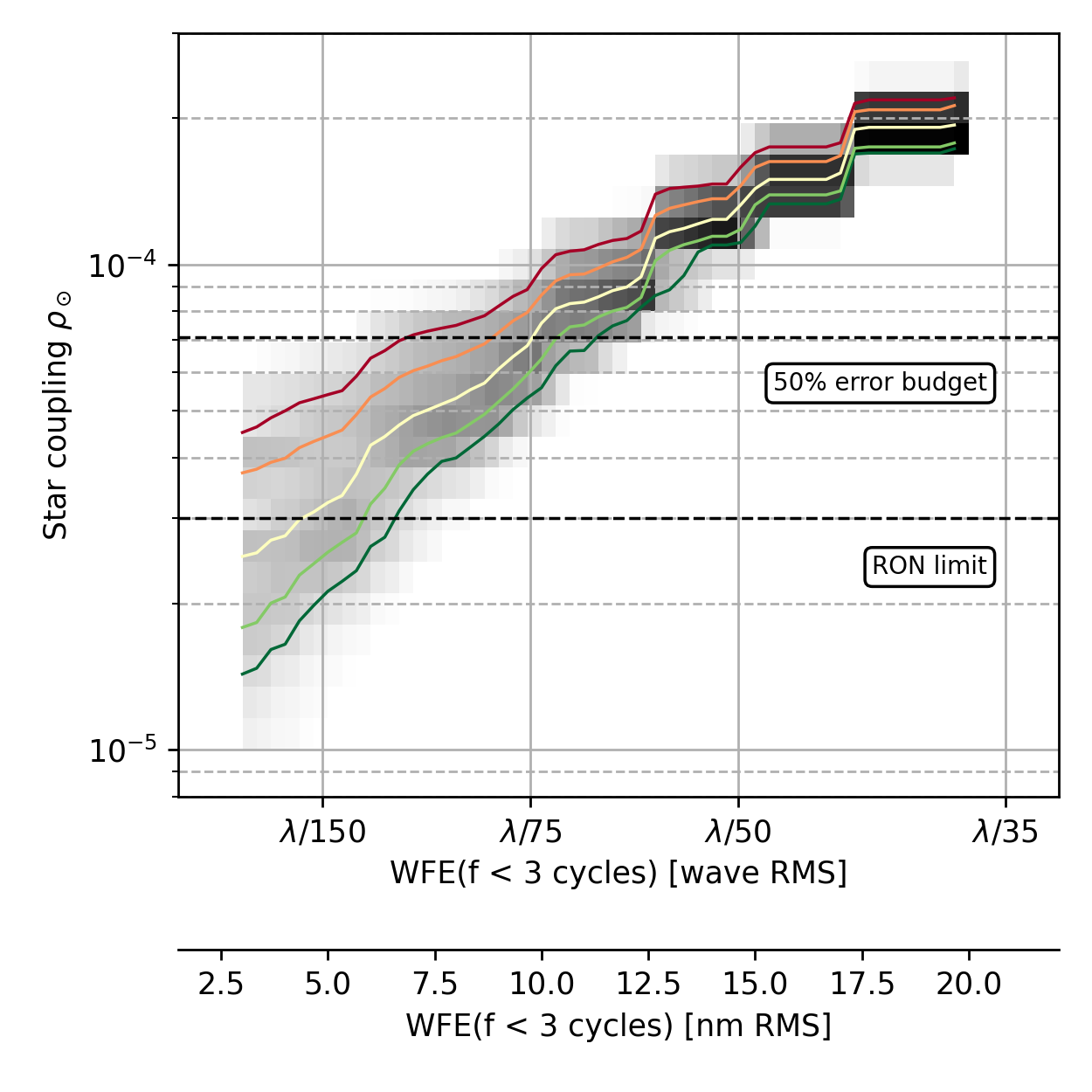}
    \caption{$\rhos$ as a function of the XAO residual error, integrated over 3 cycles.  Left: Each point resulting from a different simulation, color coded by Strehl at $\lambda$=750nm. Middle: PSD of all data binned by WFE, shown as a gray map. Green to red curves represent 10, 25, 50, 75, and 90\% percentile of best cases. Right: same as middle, filtered for Strehl$\ge$70\%.}
    \label{fig:xao_contrast}
\end{figure*}

\subsubsection{Manufacturing errors}
\label{sec:manufacturing_tol}

To assess feasibility, we also included manufacturing and alignment errors in the model, either in the PIAA optics, in the IFU or upstream in the Front-End. Tab.~\ref{tab:manufacturing_errors} presents the major tolerances we obtained from the analysis. 

We did not perform a study on PIAA sag errors, since they can be compensated with a DM as we present in Sect.~\ref{sec:wfc}. The DM correction is only partial, since it can only be conjugated with one of the two PIAA surfaces, but the impact is well within tolerances. Given current manufacturing capabilities, errors at frequencies beyond the DM correction are so small that they have a marginal impact, mostly on the Strehl/transmission level.

The tabulated values were obtained considering each term independently. To assess the effective performance of a complete system, we also performed a series of Monte Carlo simulations. The values considered are presented in the "MCMC" column of Tab.~\ref{tab:manufacturing_errors}. Note that we ignored some parameters since we have means to keep them well under control during manufacturing or alignment. This includes {\it 1)} the PIAA thickness for a rod design, which can be measured prior to making the aspheres, allowing to adapt the sag; {\it 2)} terms related to pupil chromatism or pupil size, for which a mask can redefine the pupil at the PIAA level and 'hide' the defects. As explained before, we do not consider PIAAN sag errors, so the system considered in this analysis is free of wavefront errors from optics. Results are presented in Fig.~\ref{fig:mcmc-tolerance} in the form of probability density function for each wavelength. With the tolerances considered, we have an excellent probability of obtaining a PIAAN that delivers $\rhosm \sim 2\cdot10^{-5}$ in the lab. We also investigated the performance if we consider a perfect IFU (Fig.~\ref{fig:mcmc-tolerance}, right), and see that the quality will increase drastically, very likely reaching the design value (ignoring PIAA sag errors). Combined with proper wavefront control, such a PIAAN would probably reach the foreseen contrast levels.

\subsubsection{XAO residuals}
\label{sec:wfe_tol}
XAO residuals are included in the dynamic term of our error budget. Although sensitivity to tip-tilt and coma (or even eigen modes; Sect.~\ref{sec:wfc}) give a good idea of the tolerance of PIAAN, we need a more general tolerance specification on wavefront error to design the XAO system, which must deal with all possible spatial frequencies. Given the non-linear behavior of $\rhos$ around the design position and the Kolmogorov statistics of atmospheric turbulence, such a tolerance is not straightforward to extract, so we use a statistical approach. We use the analytical model of OOMAO \citep{conan_2014a} to run several hundreds XAO simulations for a VLT telescope. We varied the parameters of the XAO like the deformable mirror actuator count ($N_{act}=40-60$) and the loop speed ($f=1-4$kHz), and the seeing conditions ($\theta=0.4-1.5$", $v_{wind}=10$m/s) to vary the loop performance and get a broad range of Strehl and low order residuals. Given the small working angle, we only consider a pyramid WFS for its sensitivity. 

From each simulation, we generate a set of statistically independent residual phase screens, for which we compute the WFE integrated over 3 cycles (which corresponds to the external diameter of our spaxels). This corresponds to about 30 Karhunen-Loeve modes on a DM. The screens are then injected in the PIAAN physical model, to compute $\rhos (WFE(f\le3\:\mathrm{cycles}) )$. 

All simulation points are presented in Fig.~\ref{fig:xao_contrast} (left), color coded by Strehl ratio. We observe that Strehl (i.e. WFE over all frequencies) correlate moderately well with low order WFE, illustrating the difficulty of establishing a simple, yet robust, tolerance. A finer analysis of this set of simulations can help define the minimum requirements on the XAO design, which is postponed to a dedicated study. 

In the middle plot, we aggregated the data in the form of probability density functions of $\rhos$ binned by low order WFE. The 5 curves present the 10/25/50/75/90 percentile of best performance. The right plot is a filtered version, only including simulations of XAO systems and seeing reaching Strehl $\ge$ 70\% at 750\:nm. This filters mostly simulations with the worst contrasts, so that median performance does not vary significantly. The analysis could also be refined considering a real seeing statistics, or even a real set of osbervations. The difference between the two distributions however suggests we can define a generic requirement based on this statistics.

It shows that our dynamic specification is reached for WFE$(f\le3\:\mathrm{cycles})\le \:\lambda/100-\lambda/75$, or 8-10\:nm RMS at 750\:nm.


\subsection{Polarization aberrations}

\begin{figure*}
    \centering
    \includegraphics[width=1\linewidth]{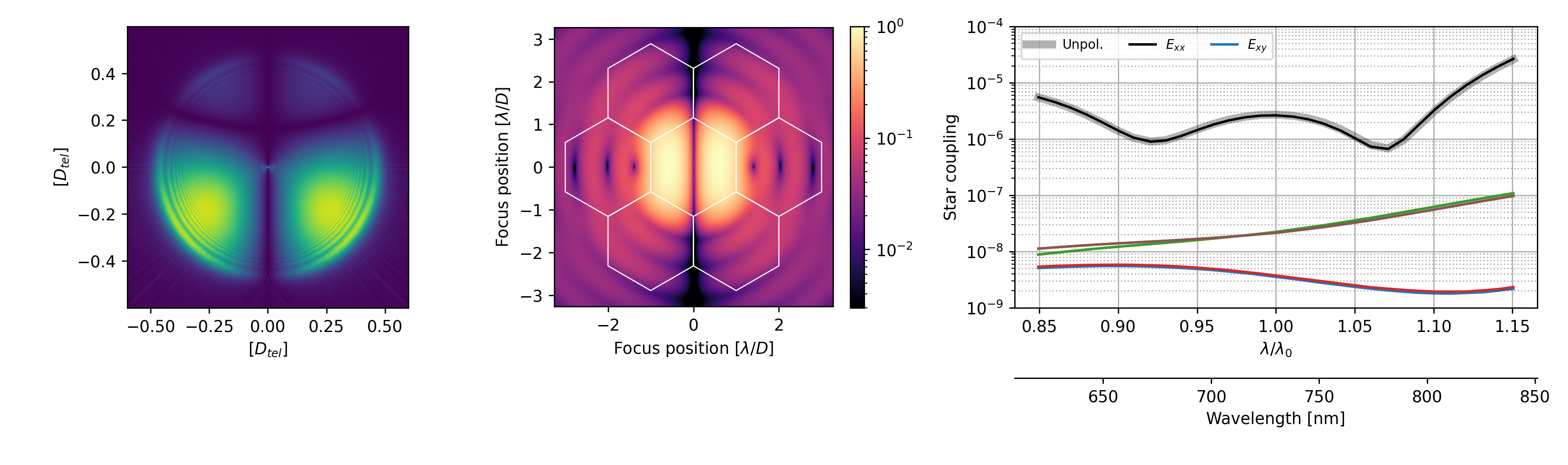}
    \caption{Example of the polarization analysis, for the crossed term $E_{xy}$ at the output of a Nasmyth fold mirror. Resulting amplitude after the PIAA (left) and the corresponding PSF (middle). Right: stellar coupling, with contribution of the 2 cross polarization to the PIAAN performance over the 6 fibers in color (some overlapping due to symmetry). The 2 lowest curves correspond to the fibers in the horizontal direction.}
    \label{fig:polarization}
\end{figure*}

Polarization dependent wavefront aberrations are due to asymmetries in mirror reflections and are a known limitation for high contrast instruments \citep{breckinridge_2015a}. Variations in the angle of incidence across reflective surfaces (concave or convex mirror, fold mirror in converging beam) lead to spatially dependent changes in electric field amplitude (diattenuation) and phase (retardance), locally altering the state of polarization (SOP). This leads to multiple independent point spread functions at the focal plane, one for each combination of parallel and cross-polarization terms. The undesired crossed polarizations also have a typical size of 2\ld due to the asymmetry in the reflection, which could be particularly harmful for such narrow working angle. In the case of unpolarized light, such effect cannot be corrected.

We use Zemax Optics Studio software to propagate linear SOP via ray tracing in a model of the VLT, including the M3 fold mirror towards the Nasmyth platform. It also includes a K-mirror derotator placed after the VLT focus in the f/15 diverging beam: it is made of 2 mirrors with angle of incidence (AOI) of 60\degree, and 1 mirror with AOI=30\degree. For a given 2D input linear polarization E$_x$ or E$_y$, Zemax computes the complex electric field in the exit pupil ($E_{xx}$, $E_{xy}$, $E_{yx}$, $E_{yy}$). Those electric fields are injected into our PIAAN model as input pupil illumination (instead of a uniform one) to assess its behavior over the 4 exit SOP. For each input state, the total coupling is the sum of both exit terms, e.g. $\rho_{x} = \rho_{xx} + \rho_{xy}$. Fig.~\ref{fig:polarization} (left \& middle) presents the resulting propagation for $E_{xy}$ alone through the PIAA optics. As expected, the core spreads slightly in the external lenslets, although it is well contained thanks to the apodization. The contrast of this PSF is much lower than from a uniform pupil (the rings have a relative intensity of $\sim 10^{-2}$), but it still contains many destructive rings throughout the lenslet. The total energy in the cross polarizations being also $\le 10^{-3}$ from the input, the contribution of the cross polarization is negligible, generally well below $10^{-7}$, i.e.\:10 to 100 times below the design value (Fig.~\ref{fig:polarization}, right).

The K-mirror rotator may also constitute an important source of polarization aberrations. 
Considering a K-mirror with a 0 or 90\degree orientation with respect to the M3, the effect between $E_{xx}$ and $E_{yy}$ is a differential tilt with respect to the unpolarized beam, with an amplitude of less than 10\;nm PV ($\le \lambda/D/60$ in the visible, or 0.25\:mas PSF offset), which is negligible in our case. In the end, the K-mirror also has a negligible impact on coupling and contrast. 

Polarization aberrations are therefore negligible for the considered contrast levels with 2\ld size lenslet.

\section{Wavefront control with the PIAAN}
\label{sec:wfc}
\begin{figure*}
    \centering
    \includegraphics[width=0.9\textwidth, trim={4.8cm 0cm 4.8cm 0cm}, clip]{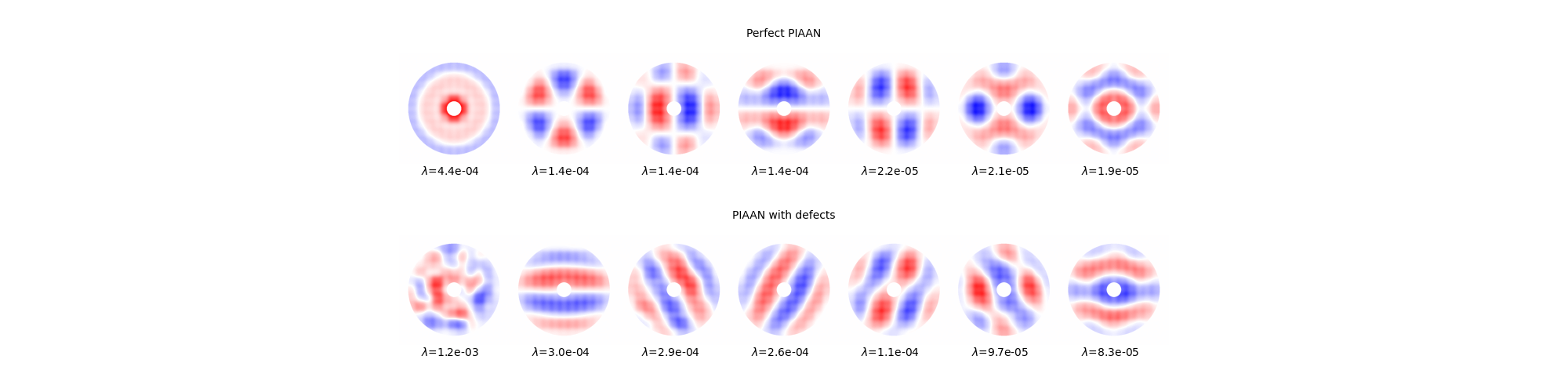}    
    \caption{Top: The 7 eigen modes obtained after calibration for a perfect PIAAN. Bottom: eigen modes in presence of defects. }
    \label{fig:eigen_modes1}
\end{figure*}

\subsection{Objective}
Considering the tight tolerances, wavefront control is fundamental to make the PIAAN operational. Not only the wavefront error requirement in itself is extremely stringent, but any defect in the PIAA optics or IFU manufacturing will shift the optimum working point from the natural design solution at "WFE=0", or even prevent it to work at desired level. A dedicated wavefront sensor near the IFU does not appear as a viable solution at such low tolerance to low order WFE, so we investigate how to perform wavefront sensing and control with the PIAAN itself. Our goal here is to "correct" static errors (PIAAN defect and static WFE) to reach $\rhosm\le 7\cdot10^{-5}$.

\subsection{Principle}

Our strategy is inspired by \citet{giveon_2007} and \citet{por_2018a}, and is also similar to the idea of Implicit Electric Field Conjugation applied to single mode fibers \citep{liberman_2024a}, extended to a fiber bundle here. We perform a calibration of the coupling variations $\delta \rho_i$ of each fiber to small input disturbance $\delta x$, with typically a couple of nanometers in amplitude. They can be injected with a DM for instance. We end up with a form of interaction matrix $M$:
\begin{equation}
    \delta \rho_i = M_{ij} \delta x_j
    \label{eq:wfc}
\end{equation}

\begin{figure}
    \centering
    \includegraphics[width=0.5\textwidth]{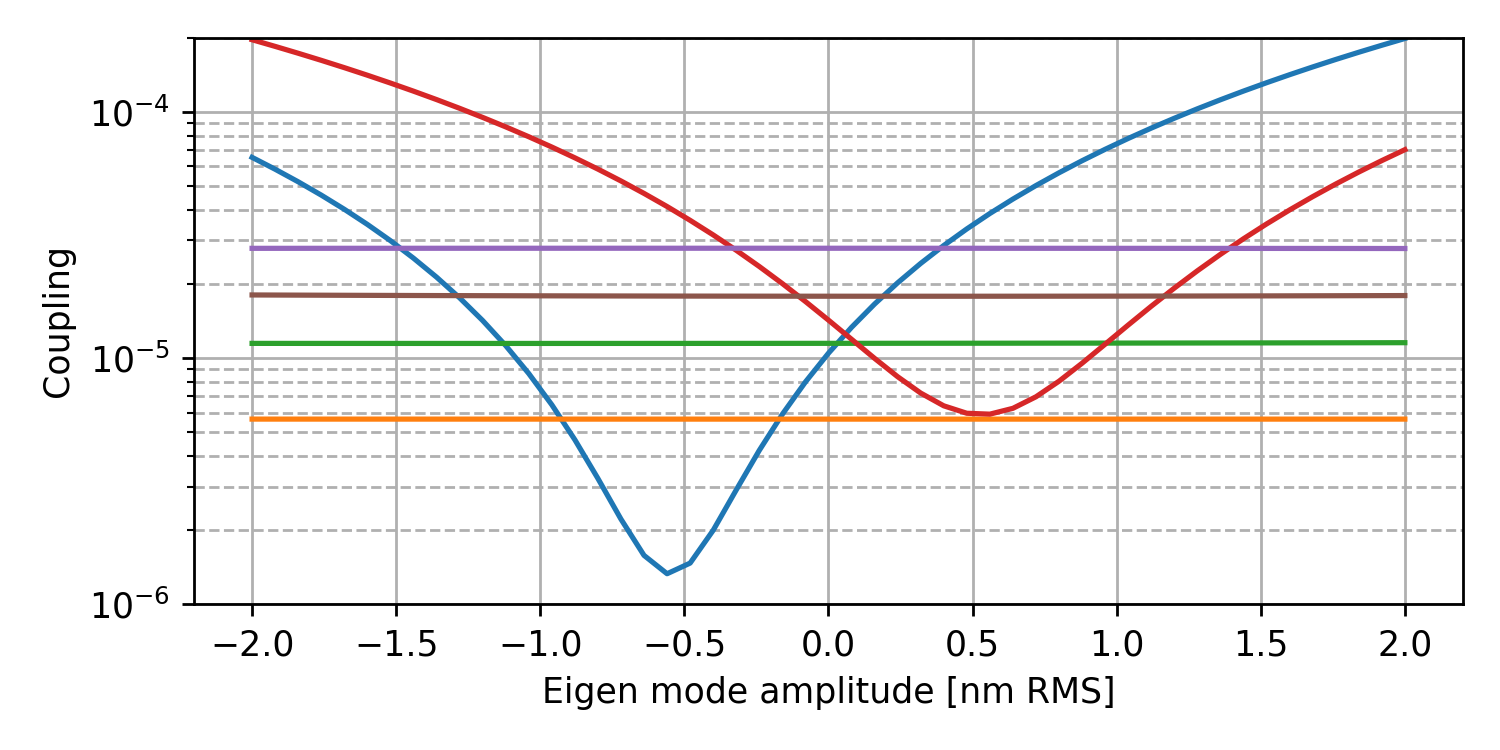}    
    \caption{Example of coupling variation over the 6 external fibers for a scan of a given eigen mode, to which 2 fibers are sensitive.}
    \label{fig:eigen_modes2}
\end{figure}

The eigen mode decomposition of $M$ leads to a series of 7 modes (Fig.~\ref{fig:eigen_modes1}, top), which can be used to optimize coupling on each fiber more or less independently (Fig.~\ref{fig:eigen_modes2}). If we include the central fiber in the analysis, the 1$^{st}$ mode corresponds to the WFE that will maximize on-axis coupling. The 6 others modes will allow control and optimization of contrast over the 6 external fibers. For a perfect PIAAN, only 3 of those modes have a real influence (3 high eigen values) due to the symmetry: one cannot change the contrast on one fiber without acting on the opposite one. As long as we deal with pupil phase errors, such modes allow to correct the contrast to its design value. Those modes can actually be decomposed as a sum of Fourier modes with frequency of about 2 cycles, and orientations of 0\degree, 60\degree, and 120\degree, corresponding to the lenslets position. With defects, an asymmetry appears and this behavior becomes clearer: Fourier modes shift slightly in frequency and phase over the pupil (Fig.~\ref{fig:eigen_modes1}, bottom) and act more and more onto individual fibers, since their variations tend to decorrelate.
By construction the eigen modes are the least tolerant modes, and we notice that our specification of $\rhos \le 7\cdot10^{-5}$ is reached with less than 1.5\:nm RMS ($\lambda/500$) variation (Fig.~\ref{fig:eigen_modes2}).

Since we measure a coupled intensity and not electric field amplitude, we cannot measure a totally linear response. The analysis is therefore only valid locally near the calibration point. If the probe is too large, we actually risk to probe through a local maximum or minimum of coupling, invalidating the linearity hypothesis of the method, and estimating modes inaccurately. So the calibration step size must be small enough, of the order of a couple of nanometers, and the procedure iterative to get modes as accurate as possible. Performing the probe with a DM upstream from the PIAA, we also end up with a set of calibrated commands, accounting for its non uniform (or even time varying) response through that optics. Since the method is by nature sensitive to only 7 low order modes, it remains difficult to solely rely on it for a global wavefront error compensation. The method is in particular flawed if the IFU is not perfectly centered on the PIAA optical axis, since applying coma can compensate for a big part of the off-axis aberration. The method therefore applies once the system is perfectly aligned and the Strehl ratio at the IFU level already high.

Note also there are different ways to build and use $M$:
\begin{itemize}
    \item A polychromatic analysis can be performed by stacking more measurements in $\delta \rho_i^\lambda$ for a given probe. More modes could also be reconstructed in that way: every test wavelength will be sensitive to slightly different spatial frequencies, bringing some diversity to the analysis, and allowing to take full benefit of higher density DM. Ideally, the eigen values should present 3 dominant terms, indicating that the contrast can be optimized by pairs of fibers, and independently of wavelength.
    \item The performance of the PIAAN assumes a perfectly rotational symmetric aperture. Any deviations, like telescope spiders, masked dead actuators, and IFU defects change this figure, so that the method can be also used to find the optimal working point and balance performance between the 6 fibers. The very limited size of such deviations could nevertheless require higher spatial frequencies than delivered by a low order DM.
    \item Considering the IFU must be rotated to fill holes in its azimuthal transmission (Fig.~\ref{fig:piaan_throughput}), we can either estimate modes for different orientation, or stack the response at different orientations, to obtain circularized modes, independent from the IFU orientation.
    \item We can finally perform the analysis on a subset of fibers. If the companion position is known, we can choose the best fiber to use and determine the one mode of interest. The complete system can be optimized for it, including the XAO, with a dedicated controller optimized to reduce lag and vibrations for instance.
\end{itemize}
We only consider here a pupil plane DM, letting us sensitive to Talbot effects, and limiting our capability to fully correct aberrations and reaching the design value.

\subsection{Procedure}
\label{sec:wfc_procedure}

We test the method in simulation with an 11x11 DM, which can in principle correct spatial frequencies beyond the field of view of the PIAAN. We also consider that the IFU is perfectly aligned to the PIAA optical axis. This is a prerequisite, otherwise optimizing Strehl through an SMF will inject tilt and coma on the DM to improve coupling.

As a first sanity check, we applied a known aberration in the DM plane (either defined by hand or following an $f^{-2}$ power law), and run an eye-doctor technique with 50 Zernike modes: the injected aberration is fully recovered after 2 iterations by optimizing coupling (i.e. Strehl) on the central fiber, with a couple of nanometers of reconstruction error. We then use the proposed calibration method using the 1$^{st}$ eigen mode: after 1 calibration of the DM and 1 scan of the eigen mode, the phase is recovered with very minor difference with the Zernike modes. 

\begin{figure}
    \centering
    \includegraphics[width=0.5\textwidth]{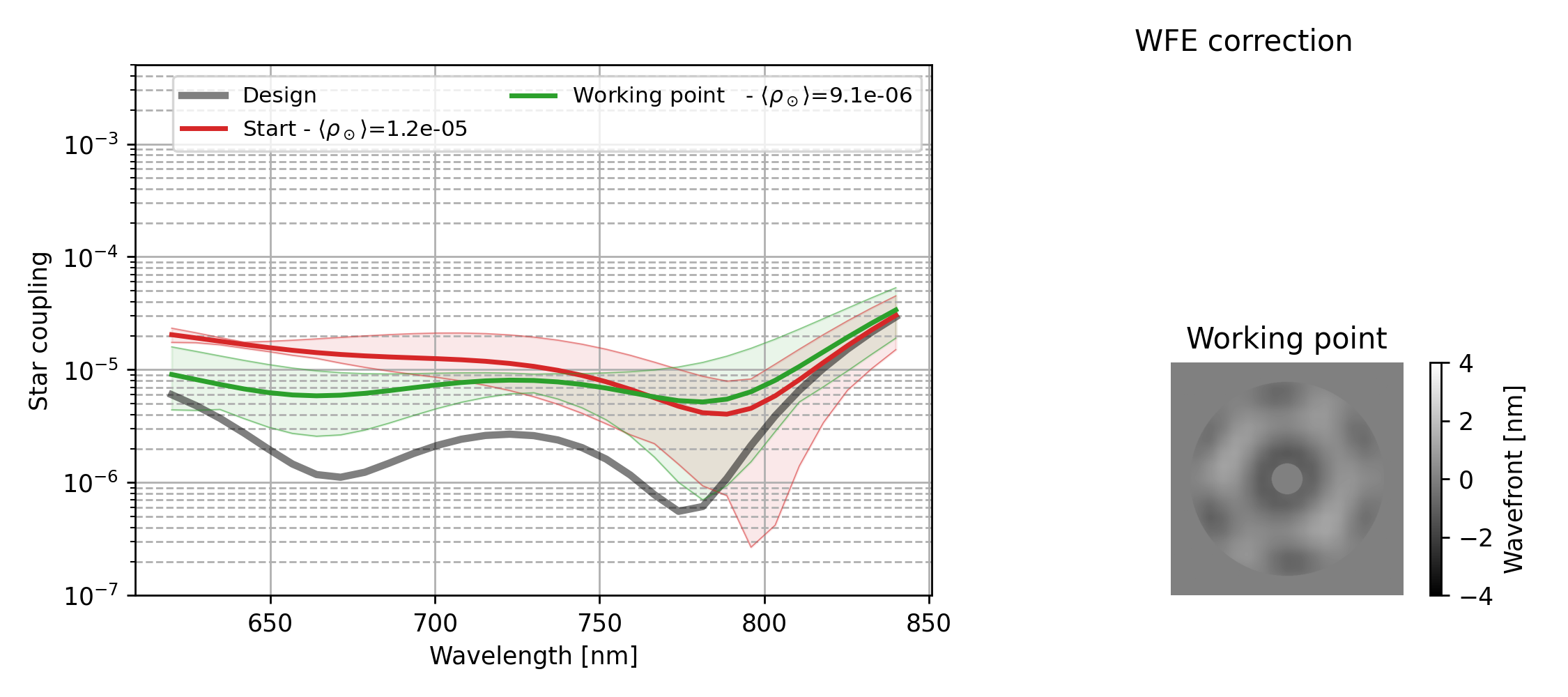}
    \includegraphics[width=0.5\textwidth]{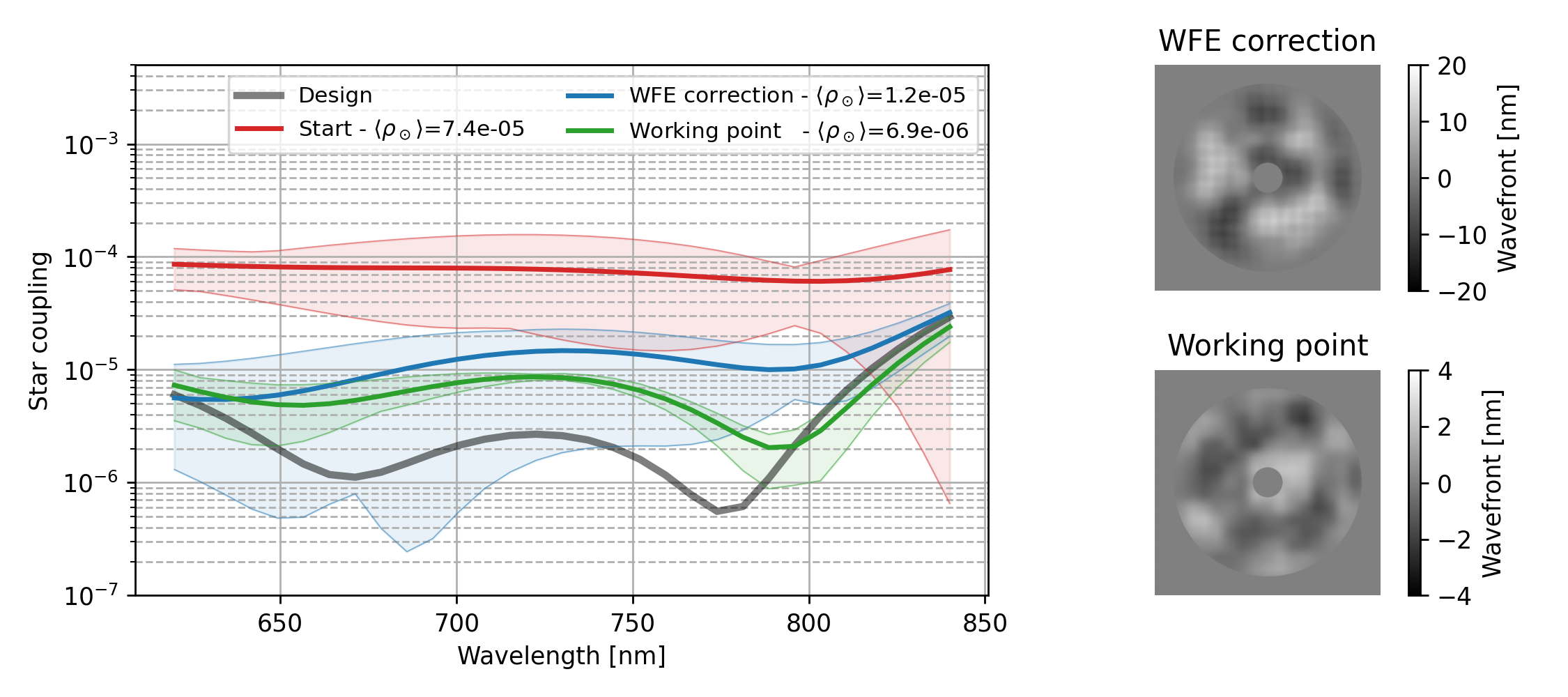}
    \includegraphics[width=0.5\textwidth]{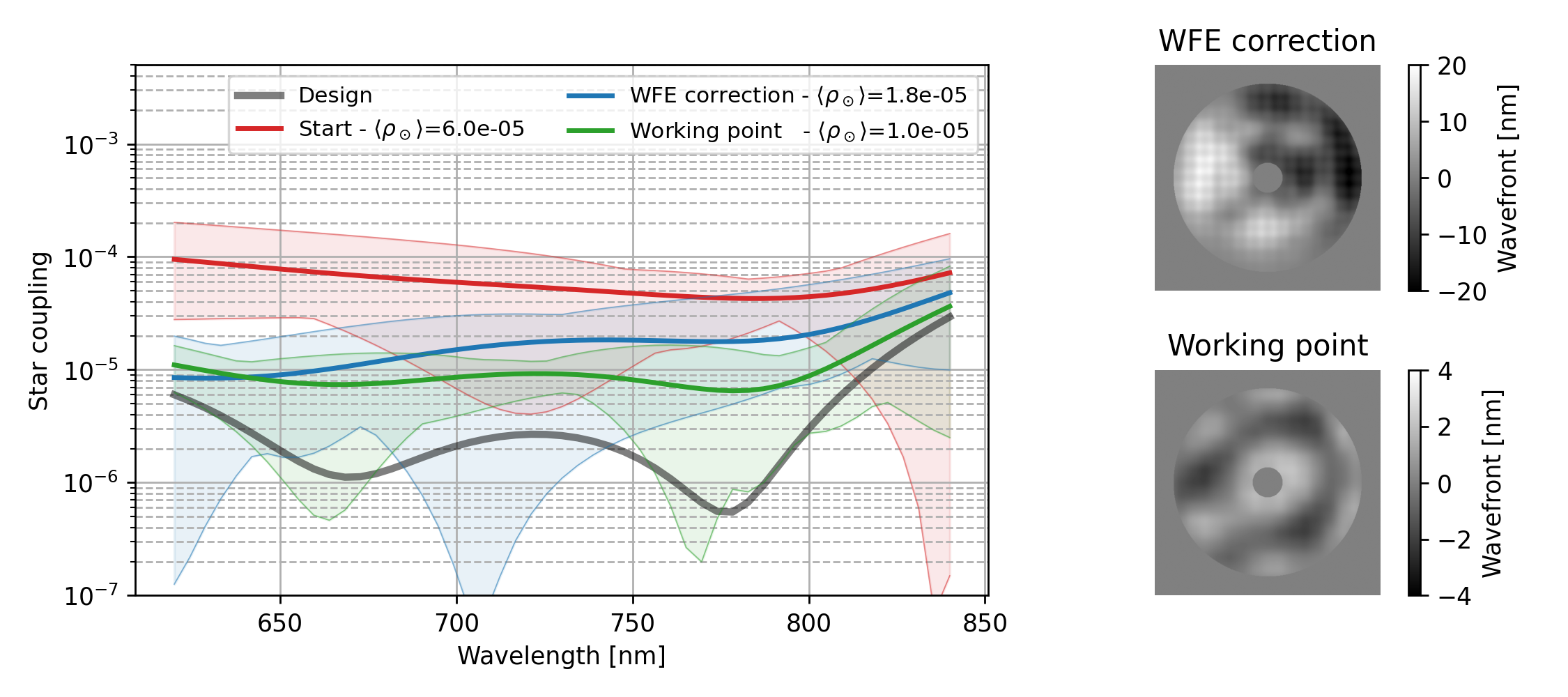}
    \caption{Result of wavefront control with IFU error only (top), PIAA errors only (middle), and the full PIAAN (bottom). Red, blue and green curves correspond respectively to the start point (DM flat), the result after WFE optimization and after contrast optimization. The filled area corresponds to the envelope covered by the 6 fibers, and the solid line to their average. The resulting DM shaped applied after each step is shown on the right of the plots. $M$ is calibrated at 730\:nm.}
    \label{fig:wfc_sim}
\end{figure}

We now add complexity with manufacturing and alignment errors on the PIAA and IFU, considering the values already presented in Tab.~\ref{tab:manufacturing_errors} for the Monte-Carlo simulations. We also add 30\:nm RMS of sag error on both PIAA surfaces, which we will try to compensate. We proceed in two steps:
\begin{itemize}
    \item {\bf Step \#1 - Wavefront correction -} We calibrate $M$ a first time and optimize coupling on the central fiber with the 1$^{st}$ eigen mode, as described before. This already recovers a big part of the performance, in particular the PIAA errors (Fig.~\ref{fig:wfc_sim}, middle). We skip this step if we consider the IFU alone (Fig.~\ref{fig:wfc_sim}, top).
    \item {\bf Step \#2 - Contrast optimization -} Due to defects in the IFU and Fresnel propagation in PIAA (Talbot effects), $\rhos$ is not yet optimal and it presents strong inhomogeneities with wavelength and between fibers. We recalibrate $M$ with only the 6 external fibers to optimize $\rhos$. For each eigen mode, we perform small scans around the calibration point (similar to a standard eye-doctor procedure), with amplitude between 1 and 5\:nm RMS. Since we do not know the optimal value of $\rhos$ we want and can reach, this scan is necessary, we cannot simply inverse M to reach a given contrast. We are not interested in obtaining extremely deep nulls at a particular wavelength: our goal is instead to obtain a balanced system, so that SNR is also balanced between fibers and at all wavelengths. So, we minimize the dispersion of contrast over the 6 external fibers for each scan, instead of optimizing contrast for one particular fiber (the procedure actually does not converge easily if we proceed that way). For systems within our tolerances, 2 to 3 passes over the 6 modes is enough to converge without an apparent ambiguity. Once close from an optimal solution, optimizing two opposite fibers at a time also appears more effective. If during an iteration, the working point moved too much, the fibers get naturally less independent for a given eigen mode and an intermediate calibration is performed. 
\end{itemize}
We demonstrate this procedure with PIAA or IFU errors alone, and finally the complete PIAAN (Fig.~\ref{fig:wfc_sim}), with a common set of errors to evaluate the impact of each component. We perform a monochromatic optimization at 730\:nm. With sag errors, $\rhos$ starts at $\sim 10^{-4}$, with variations by a factor 10 between wavelengths and fibers. The DM is particularly effective to compensate the PIAA errors and to balance contrast. It is a bit less effective with the IFU, the far-field of the SMF being conjugated to the focal plane. Introducing only fiber core misalignment leads to better results, with almost complete recovery of the design performance. This is more difficult with far-field alignment errors. We notice the benefit of the PIAA apodization, which allows the contrast optimization at no cost on transmission with a wavefront offset below 2\:nm RMS.

All errors considered, we seem to reach a limit at about $1\cdot10^{-5}$ (also observed with different sets of errors). This is probably due to off-pupil errors (Talbot effects and IFU errors) that we attempt to correct with a single pupil DM. Without too much surprise, doubling the actuator density does not change much the eigen modes nor improve the end results. 

The achieved contrasts are about a factor 7 better than our requirements, so that performance should be ultimately limited by XAO performance (see Sect.~\ref{sec:wfe_tol} and Fig.~\ref{fig:xao_contrast}), so we do not study the benefit of a second DM. 

It is worth noting that this method could be applied to any singlemode fiber coronagraphic IFUs.

\subsection{Non Common Path Aberrations in operations}

In practice, we will want to find the working point of the PIAAN during integration or daily calibrations. The working point, limited by the PIAAN alignment and defects, should be stable in time. Optical aberrations are more prone to evolve over time scale of 1\:hour and with amplitudes significant for small inner angle coronagraphs \citep{vigan_2022a}. During the night, NCPA could be calibrated on an internal light source: with a camera working at several 100\:Hz (or even photodiodes much faster than any DM settling time), and fast flip mirrors and shutter, such a calibration could be performed in a couple of seconds during an hour long exposure.

\section{Prototyping and test results}
\label{sec:prototyping}

\subsection{Integral Field Unit}
\label{sec:ifu}

The IFU unit consists in a bundle of 7 standard step-index SMF, packed in an hexagonal array. A micro-lens array is then 3D-printed on top of the bundle. We thoroughly characterized their performance to eventually adapt the microlens design, as well as to estimate the readiness level of the novel 2-photon polymerization 3D technique for astronomical use.

\subsubsection{The singlemode fibers}
\label{sec:smf630hp}

\begin{figure}
    \centering
    \includegraphics[width=0.48\textwidth]{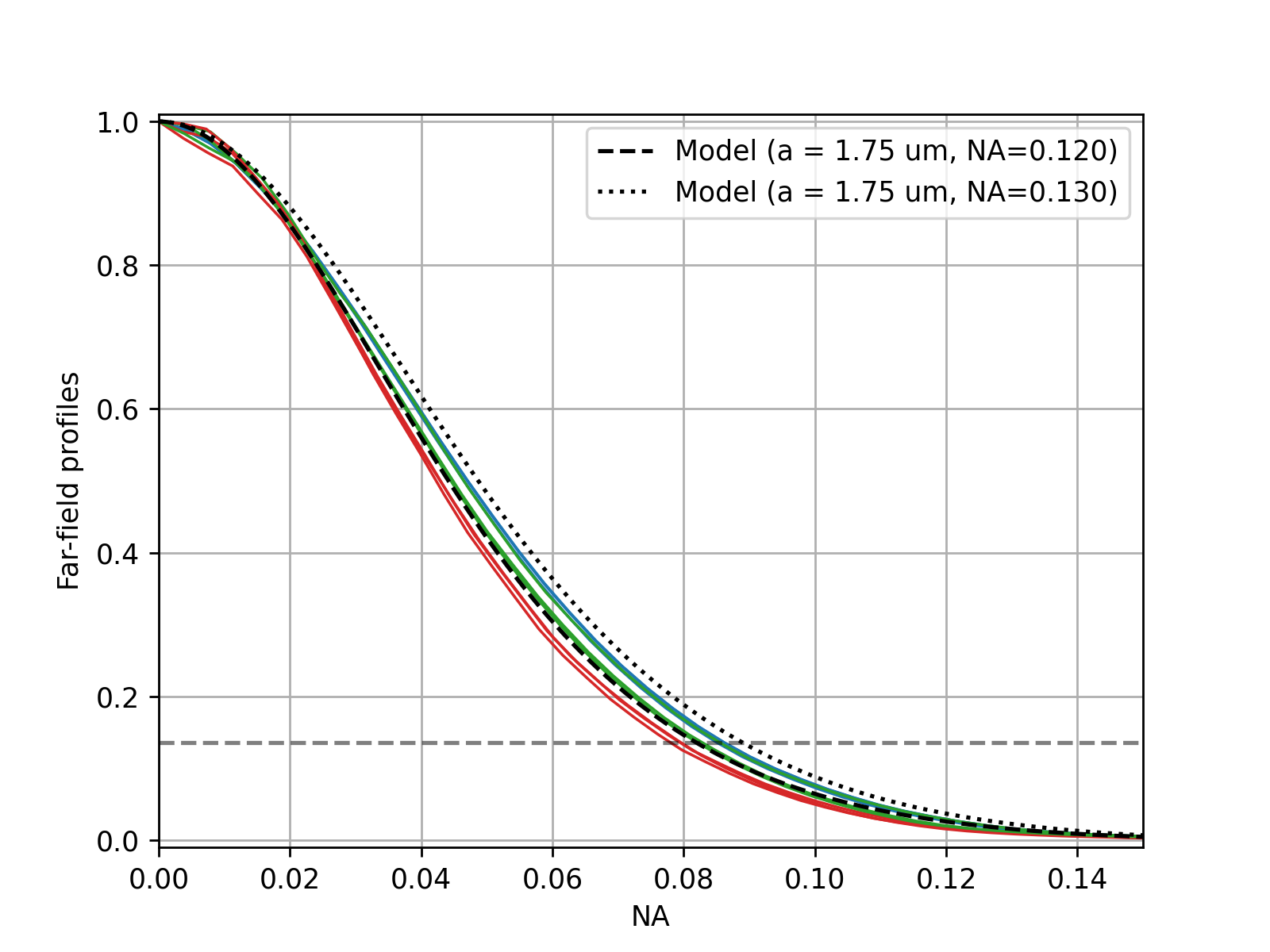}
    \caption{Thorlabs SM630HP NA characterization for different fibers (color lines). SMF models in dashed and dotted black lines.}
    \label{fig:smf_charac}
\end{figure}

We use off-the-shelf Thorlabs SM630HP singlemode fibers with a core diameter of 3.5$\mu$m and NA=0.13. They were thoroughly characterized, in particular regarding the NA which must be known to better than 5\% to optimize the MLA design. 
We directly image their far-field on a CCD, without optics. We acquire images at different CCD-to-fiber distances, which allows to extract the exact distance for each step and measure the angular scale of the setup. A theoretical LP$_{01}$ mode is then fitted to the data. We measured several samples coming from Thorlabs (bare or connectorized) or a sample provided by SQS/AMS with their bundle. We estimated NA(1/$e^2$) = 0.083 ($\pm\:3\%$) which corresponds to geometrical NA=0.120 instead of the data sheet 0.130 (Fig.~\ref{fig:smf_charac}). This leads to a single-mode cut-off wavelength of 550\:nm.

One of the fiber characterised here is kept as a reference for future comparative tests, with the bundle and IFU in particular.

\subsubsection{Bundle arrays}
\label{sec:bundle}

Fiber bundles prototypes were produced by AMS/SQS with a pitch of 125$\mu m$ (Fig.~\ref{fig:ifu_proto}, A), after removing the fiber coating. For these prototypes, we use an 'octopus' configuration, with 7 FC/PC fibers as output of the hexagonal bundle. For the final bundle, the 7 fibers will be grouped in a linear slit at the output.

The two bundles are of excellent quality, with a core-to-core pitch of 125\:$\pm$\:0.7$\mu$m as estimated from a gaussian fit to the modes. Far-field co-alignment (i.e. mechanical parallelism of fibres), as well as perpendicularity to the mechanical surface, is also well below 0.3\degree\:for all fibers (see Tab.~\ref{tab:ifu_perf}).

\subsubsection{Micro-lens arrays}
\label{sec:mla}

\begin{figure*}[t]
    \centering
    \includegraphics[width=1.\textwidth,  trim={3cm 0 2cm 0}]{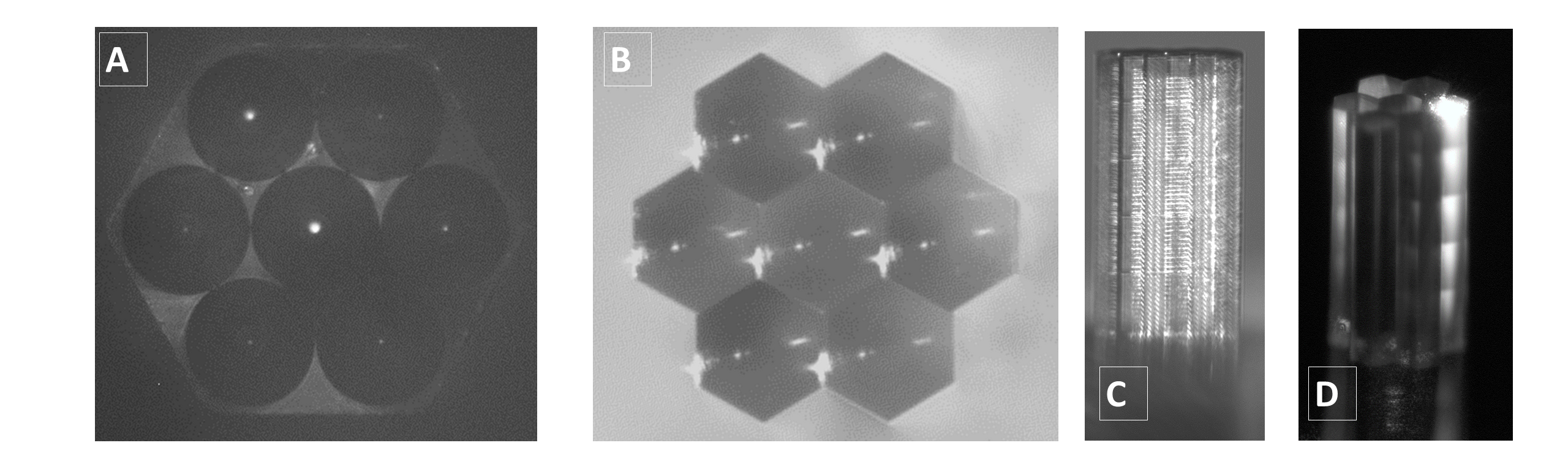}
    \caption{IFU prototype with 125$\mu$ lenslet size. A) Bundle front face; B) Same view as A) after the MLA has been printed on it; C) Side view of the MLA, with illumination from the back, showing some of the printing structure. The lenslet height is 0.7726\:mm. D) Three-quarter view showing the 7-lenslets. A laser source is injected from the fiber to the lenslet for one of them. Diffusion in the lens material shows the beam.}
    \label{fig:ifu_proto}
\end{figure*}

The micro-lens array (MLA) is made by Keystone Photonics (formerly Vanguard) using the 2-photon polymerization 3D printing technique. The technology allows maximum flexibility and accuracy to create lenses that are optimized for the bundle. The technology promises the alignment of every fiber core to their respective lenslets optical axis with an error of $\pm\:100$\:nm, well in the PIAAN specifications. A 1$^{st}$ prototype with a pitch of 250$\mu$m was tested and showed encouraging results \citep{kuhn_2022a}. The large pitch led to a very complex fabrication process, requiring high volume to print, multiple printing passes, realignment of the machine, etc. A second prototype with a minimal pitch of 125$\mu$m, matching the fiber bundle pitch, was produced with a height of 0.7726\:mm. As seen in Fig.~\ref{fig:ifu_proto} (B to D), the lenslets present some structures on the edge, and five printing layers are visible. Despite the straight pillar design, leakage and coupling from one fiber to the other was not observed at the level we are interested. The physical distance between the cores certainly helps in this. For one prototype, one MLA surface was measured by Keystone with an interferometer: the sag is extremely accurate and presents a 10\:nm annular structure on its surface, due to the printing technique. Injecting those data in a model of the MLA shows it has a negligible impact on performance, with only a few percent of light lost.

On all prototypes, we also observed a systematic global tilt of $\sim$1\degree of the fiber bundle far-field when back-illuminated onto the MLA. The reason is not clear since we measured that the far-field is perfectly perpendicular to the bundle surface (see previous section), and no such angle is observed between the bundle surface and the MLA. This surface is also used by Keystone to align the bundle into their machine. In theory, this should have a limited impact on contrast.

\subsubsection{IFU performance}
\label{sec:ifu_perf}

To characterize the IFU, we used a simple bench, the same presented in \citet{restori_2024a}. It is made of two doublet lenses working at f/100 to get a proper plate scale of 2\ld=125\:$\mu$m at the MLA level. The IFU is placed at the bench focus on a 5-axis motorized stage to optimize coupling into fibers. We back-illuminate the IFU to align it angularly to the pupil, and laterally to the input fiber. After this preliminary alignment, there is no ambiguity in optimizing the 5-axis by injecting light from the bench source to the IFU.

The complete IFU with a clear pupil mask. In this configuration, the maximum theoretical coupling amounts to $\rho_{max}^{theo} = 80\%$ at $\lambda=633\:$nm. The measurements lead to maximum coupling between 36 and 57\% (Tab.~\ref{tab:ifu_perf}). Many tests were performed to identify the origin of the loss, whether due to fibers or to the MLA (absorption or premature aging, volume grating effects, etc.), but none allowed us to obtain a consistent picture of the issue. Despite those rather disappointing coupling values, we could still obtain very high contrasts, as presented in Sect.~\ref{sec:piaan_perf}.

\begin{table*}
    \centering
    \begin{tabular}{l|ccccccc}
        \hline \hline
               & F1 & F2 & F3 & F4 & F5 & F6 & F7 \\
        \hline
        \multicolumn{8}{c}{Fiber bundle}\\
        \hline
        Core to core distance [$\mu$m]  & - & 124.0 & 124.5 & 124.5 & 125.1 & 126.1 & 124.8 \\
        Far-field angle [\degree] &  0 &0.03& 0.13 & 0.02 & 0.03& 0.16& 0.14 \\
        \hline
        \multicolumn{8}{c}{IFU}\\
        \hline
        Max.  coupling $\rho_{max}$ & 0.42  & 0.45 & 0.45 & 0.38 & 0.40 & 0.29 & 0.34 \\
        $\rho_{max}/\rho_{max}^{theo}$ & 0.52 & 0.57 & 0.57 & 0.47 & 0.50 & 0.36 & 0.43 \\
    \end{tabular}
    \vspace{0.3cm}
    \caption{Fiber bundle and IFU measured performance. F1 is the central fiber.}
    \label{tab:ifu_perf}
\end{table*}

\subsection{PIAA optics}
\label{sec:piaa}

\begin{figure*}
    \centering
 \includegraphics[width=0.9\textwidth]{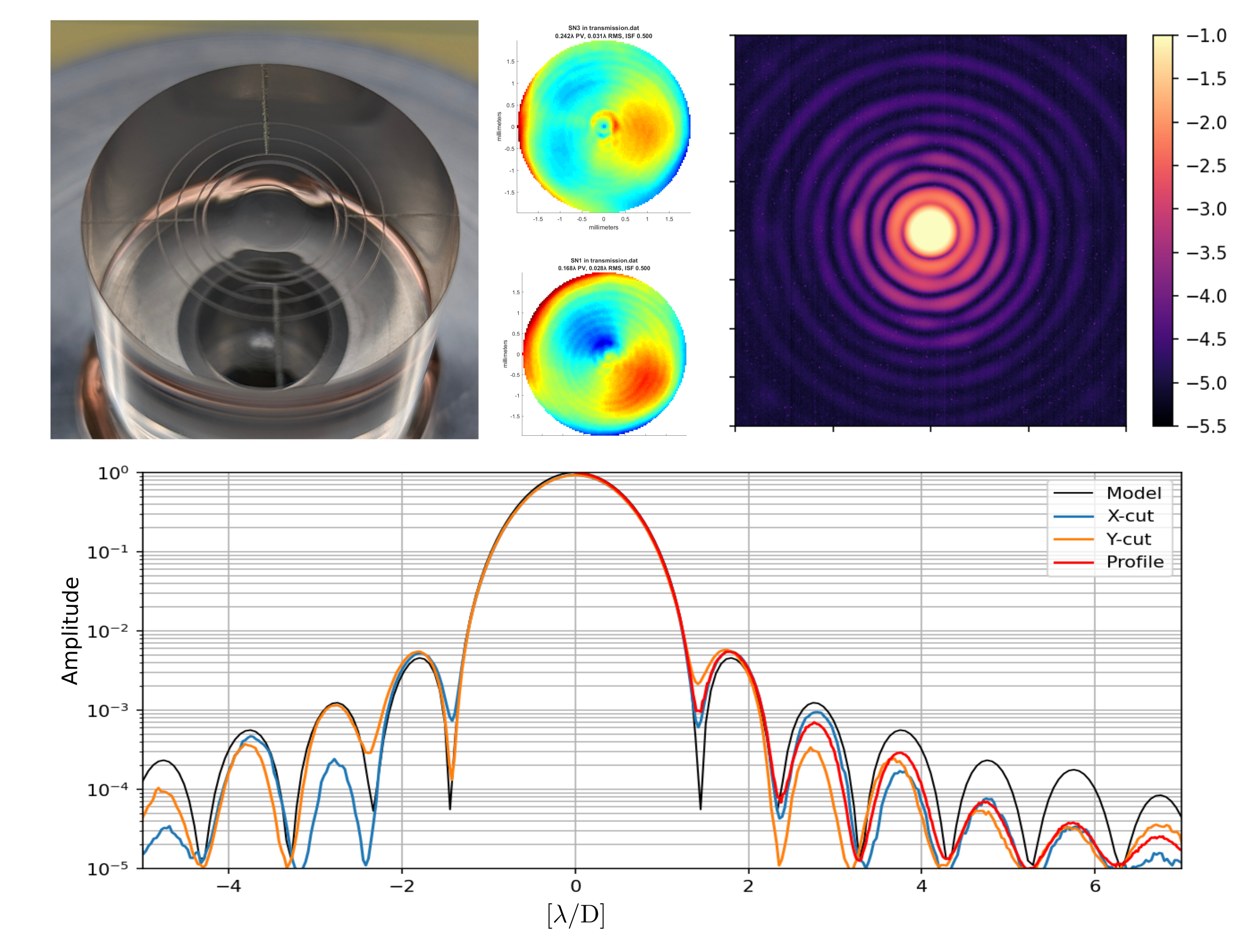}
    \caption{PIAA prototype. Top, left: Picture of one PIAA optics before coating. Top, middle: WFE error of both prototypes as measured by Nutek. Top, right: PIAA PSF in log scale at $\lambda$=633nm. Bottom: Cut views and integrated profile of compared to the theoretical PIAA PSF in black.}    
    \label{fig:piaa_proto}
\end{figure*}

Two PIAA optics were produced by Nutek, with the rod CaF$_2$ design presented in Sect.~\ref{sec:piaan}, with a pupil of 4\:mm in diameter and a glass thickness of 20\:mm (Fig.~\ref{fig:piaa_proto}). The centering of both aspheres must be kept below 30\:$\mu$m, which was well within Nutek capabilities. The thickness of the CaF$_2$ blank was measured prior to turning the optics, allowing to adjust the sag of both surfaces. The rather small sag and choice of a rod design led to an excellent optical quality, with transmitted WFE $\le$ 30\:nm RMS for both prototypes (Fig.~\ref{fig:piaa_proto}). The PSF obtained on a bench working at f/100 confirms the very high optical quality even at visible wavelength ($\lambda$=633nm). It looks slightly more apodized than expected, and not having access to a machine able to measure both sag, it is difficult to estimate its origin. This could have a negative impact on the contrast performance by changing the balance between the rings.

 
\subsection{Experimental set-up}
\label{sec:high_contrast_bench}

\subsubsection{Design}
We built a bench to characterize the PIAAN (performance, stability) and to test wavefront control strategy with a DM. The goal of this bench being a polychromatic characterization of the PIAAN, the bench is reflective. A refractive solution with off-the-shelf components f/100 beams was also studied. Chromaticity of this solution was theoretically well within our WFE specification over 30\% bandwidth, but it was unclear if produced optics can match it, which would have left us with strong doubts on PIAAN performance if they were not reaching expectations.

The bench is built exclusively from off-the-shelf components. The main path of the bench (Fig.~\ref{fig:high_contrast_bench}) consists of three replicated Off-Axis Parabola (OAP1, OAP2, OAP3) with f=279mm (Edmund Optics EO90-978 / Shimadzu 275-07-4040) working at f/70, and a fourth f=254mm 30\degree\:OAP (Thorlabs MPD2103-P01; OAP4) working at the optimal f-number of 63.5 for an IFU with 125$\mu$m lenslets. A Boston Micromachines Multi-3.5 DM (12x12 actuators) is placed between OAP1 and OAP2 to control wavefront in the bench. A chromium pupil mask and the PIAA optics are conjugated to the DM in a pupil plane between OAP3 and OAP4. The chromium mask defines the pupil and is placed about 1\:cm before the PIAA, which is within our tolerance for a 4\:mm pupil size. 
The PIAA is mounted inside a four-axis gimbal mount (Standa 5GTM1).
 The IFU is placed on a 5-axis motorized stage: three-axis translation stage (Thorlabs MTS25) to align the IFU to the PIAA optical axis, and a 2-axis motorized Gimbal mount (Zaber OMG) to align the IFU far-field to the PIAA pupil.

Post-PIAA focal and pupil imaging are provided thanks to a pickup beam splitter, followed by two cameras (CAM1 and CAM2). The beam splitter presents a small wedge dispersing the PSF at the IFU level, so a second plate with an equivalent wedge is placed after to compensate it.

To perform polychromatic measurements, a simple fiber-fed prism-based spectrometer was built (Fig.~\ref{fig:high_contrast_bench}, top). The IFU bundle is connected to another bundle, made of 7 multimode fibers with 50$\mu$m core, rearranged in a slit. The choice of multimode fibers was made to limit chromatic coupling effect at the connector level, that could bias the characterization. The spectrograph resolution is limited by the fiber size to 100. The dispersed fiber signal naturally presents a high modal noise, but since we are interested only by information at a resolution of a few 10, it averages well and we do not plan to add a mechanical scrambling device. Due to the limited dynamic of the spectrograph camera, the central fiber is attenuated by putting both the central IFU fiber and central slit fiber $\sim$20\:mm apart, leading to an achromatic and very stable attenuation factor of 400. Measurement over a dark area of the sensor shows that contrast is limited to $10^{-7}$ in a single exposure. The wavelength scale is calibrated by using 3 laser sources at $\lambda$=638, 730 \& 850\:nm.

The full bench and spectrograph are placed under an enclosure to eliminate turbulence and stray light.

\subsubsection{Alignment and characterizations}

Optical alignment based on the Optics Studio and a Solidworks modeling proved very reliable. The high f-number is very forgiving, leading to tolerance on the OAPs alignment of the order of $\pm$\:2\:mm and $\pm$\:1\degree\: regarding WFE. The final f-number is accurate to 0.5\% after measurement of the PSF at focus.

The pupil of the system is defined by the chromium mask positioned just in front of the PIAA. Different pupils are available: a 4\:mm pupil without central obstruction and a 4\:mm VLT pupil without the spiders, and variations of these two pupils with a slightly oversized or undersized primary and secondary. This resulted in aperture diameter varying from 3.6\:mm to 4.4\:mm in 0.08\:mm steps, and in secondary obstructions with a diameter varying from 0.64\:mm to 0.78\:mm in 0.03\:mm steps, step size being related to the tolerance on the pupil magnification. The mask was patterned using photolithography with smallest details of 1\:$\mu$m. Note that the PIAA optics has been designed to reduce the VLT obstruction of 16\%: it however appears the secondary obstruction in the mask has actually no impact on the performance, as long as it is smaller than the one considered for the PIAA design. 

\begin{figure*}
    \centering
    \includegraphics[width=1\textwidth]{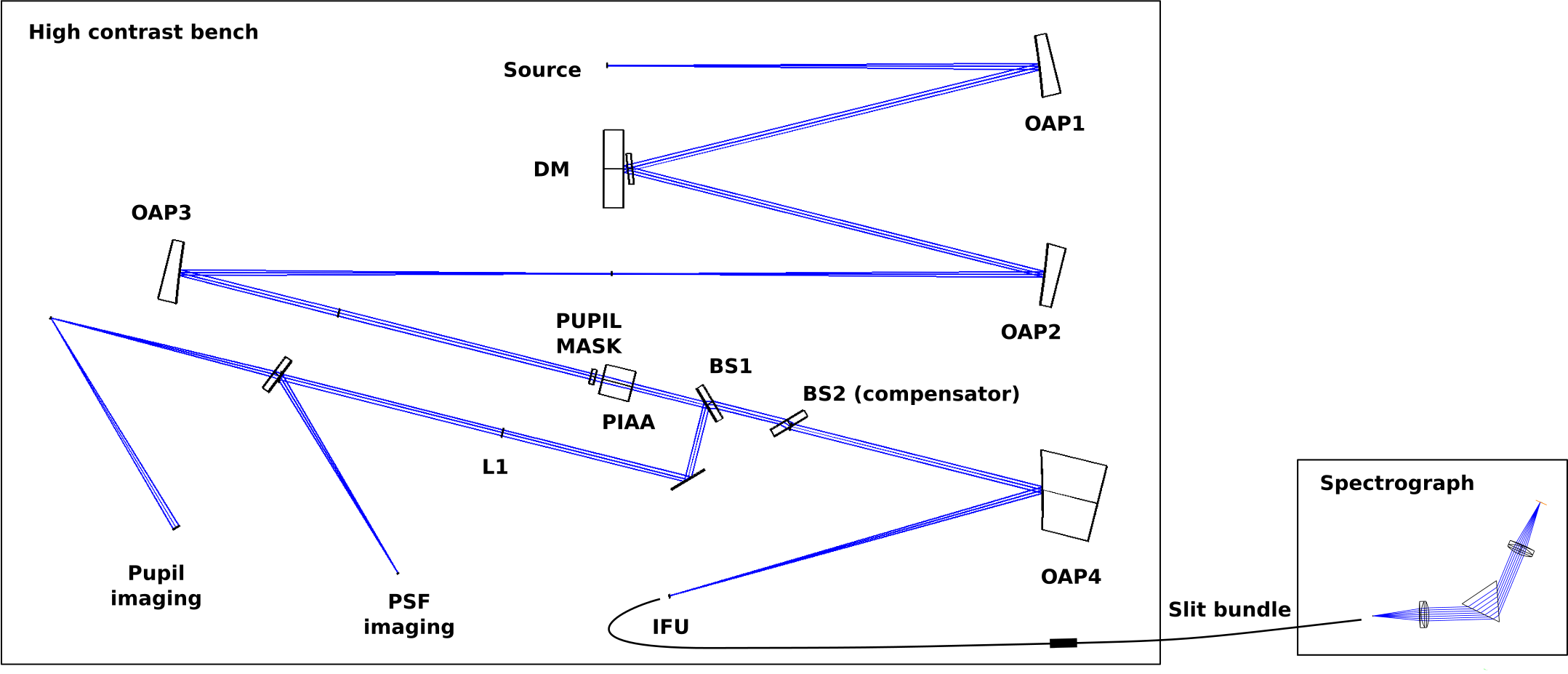}
    \includegraphics[trim={1.5cm 0 1.5cm 0}, clip, width=1\textwidth]{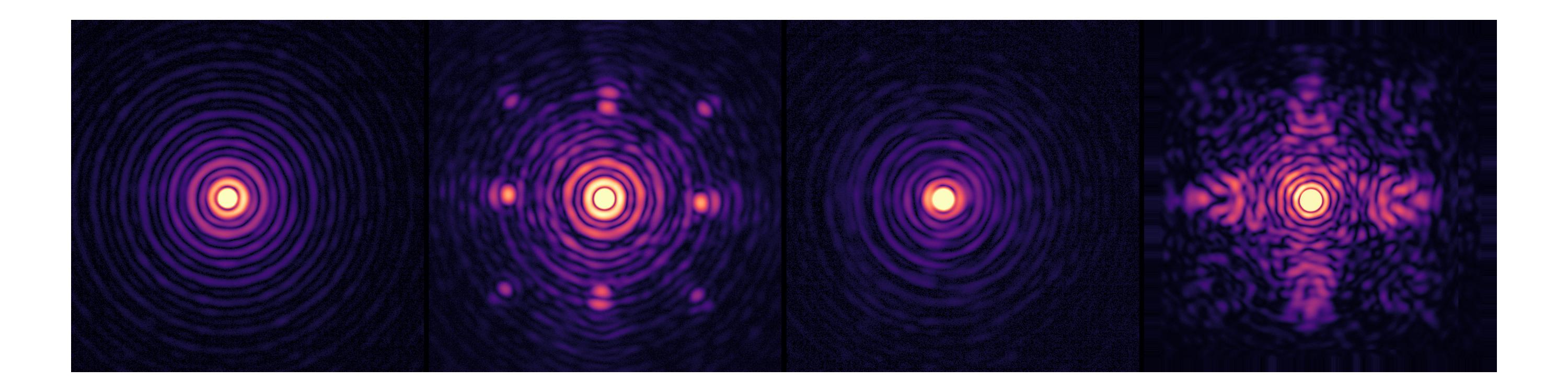}
    \caption{Top: Optics Studio design of the RISTRETTO high contrast bench and spectrograph. Bottom: PSFs at the IFU level at $\lambda=633\:$nm. The two on the left are obtained with a VLT pupil, with a flat mirror in place of the DM for preliminary alignment, and with DM optimized in a best flat position. Ditto on the right two, after aligning the PIAA, also with flat mirror and with best flat DM.}
    \label{fig:high_contrast_bench}
\end{figure*}

The deformable mirror was characterized using an interferometer with a 5\:mm beam. The response matrix of the DM was computed from this calibration, which was then used to computed a best flat command set, leading to a final surface error of 18\:nm RMS, dominated by the DM actuator structure at 12 cycles. 

The bench was first aligned with a flat mirror in place of the DM, and shows an excellent optical quality (Fig.~\ref{fig:high_contrast_bench}, bottom) without and with the PIAA optics. The optical quality of this reflective bench is lower than our original refractive IFU test bench, designed at f/100 \citep{restori_2024a}, and also used to obtain the PIAA PSF of Fig.~\ref{fig:piaa_proto}. The quality is actually limited by the last diamond turned OAP4, which presents a roughness of 10\:nm RMS. At the IFU focus, we do not observe ghosts from the usual grating structure resulting from the diamond turned manufacturing. The PSF measured at the IFU focus is nevertheless clearly of lower optical quality than the one observed on the imaging arm, where only lenses are used: we observe twice more diffraction rings on the latter and integrated PSF profiles defer at $\sim 10^{-6}$ level. 

The DM was aligned visually to the PIAA mask by pulling four actuators in a square shape around the secondary obstruction. The pupil camera was then used to center the resulting structure to the pupil mask. Since the pupil is smaller than the DM active surface and the wavefront control procedure calibrate the DM influence on coupling, an accurate centering is not required. 

To maximize Strehl ratio, we used then an eye-doctor technique, using a Zernike basis or the DM eigen basis. Both methods showed similar results. We obtained best results when normalizing the PSF to the energy contained within the correction area of the DM. While the PIAA PSF is improved, the diffraction rings still show some roughness, and in the case of the second and third rings a slight asymmetry. We can observe the strong off-axis aberration of the PIAA to the satellite spots at 5\ld, generated by the DM structure (as well as much fainter at 7.5\ld). The DM best flat measurement was injected into the PIAAN model, degrading null to $\rhosm \sim 2\cdot10^{-5}$. This can be considered as the performance limit of this bench since those structures cannot be corrected by the DM itself.

Those spots cannot be removed with a spatial filter at an intermediate focus, since it would filter the pupil for which the PIAA has been designed. A DM with a higher number of actuators might be required to push the satellites further in a future implementation.

\begin{figure*}
    \centering
    \includegraphics[width=1\linewidth]{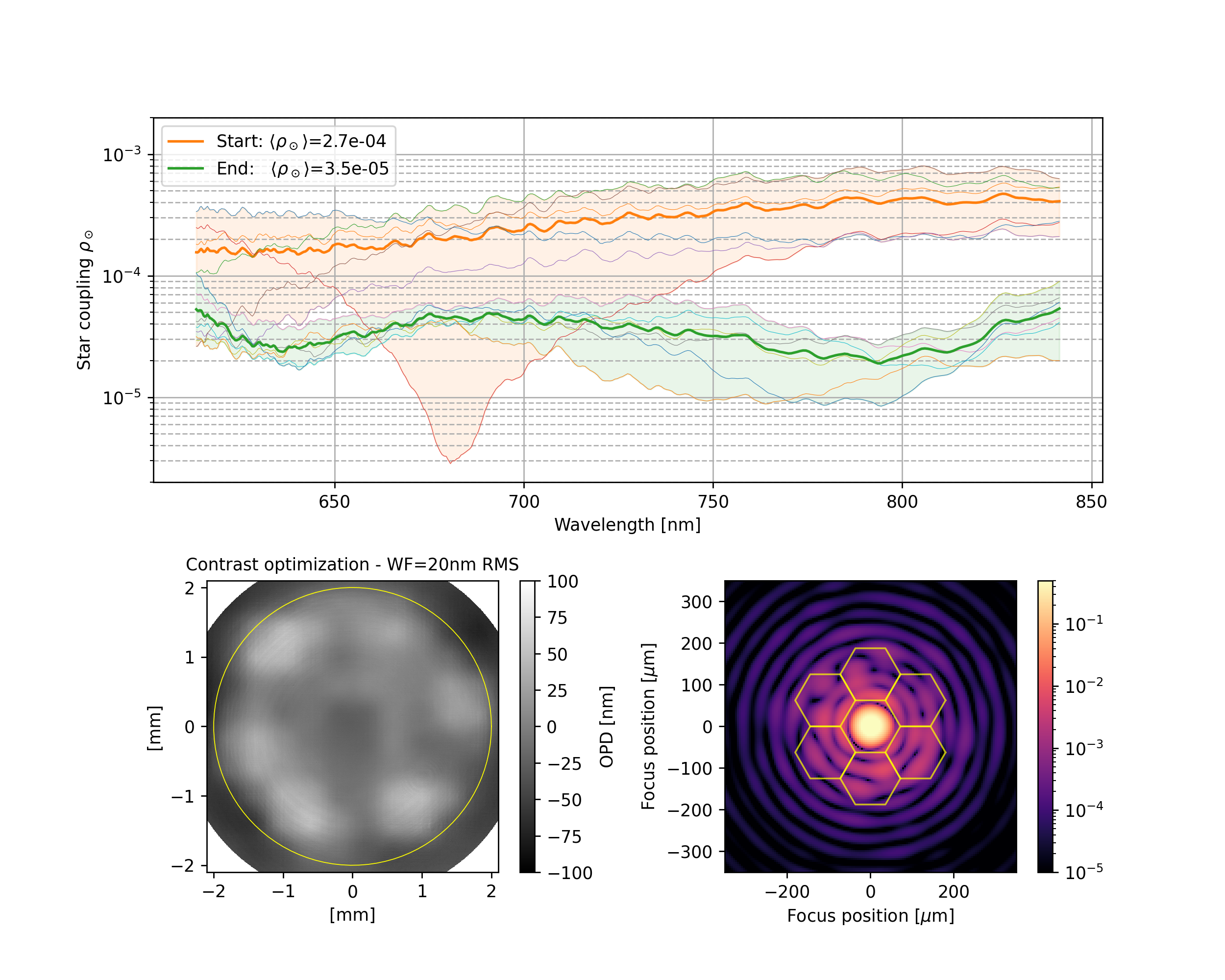}
    \caption{Top: Result of the contrast optimization of the PIAAN. The starting point is in orange, after Strehl optimization. The filled area shows the envelope covered by the 6 spaxels, each shown by a colored curve in the envelope. The end result is displayed in green, also with the curve of the 6 spaxels and the corresponding envelope in green. Bottom left: working point OPD applied on the DM. The yellow circle represents the 4\:mm pupil size of the PIAA. Bottom right: resulting PSF at $\lambda$=633\:nm.}
    \label{fig:piaan_optim}
\end{figure*}

\subsection{PIAAN performance and wavefront control}
\label{sec:piaan_perf}

We finally put our full prototype and optimization procedure at test. Flux is measured on the spectrograph for each spaxel on-axis $F_\odot$, and off-axis $F_\oplus$, by adding a 2\ld\:tilt on the DM to center the PSF on the desired spaxel. $\rhos$ is estimated by considering the optimal value of $\rhop = 0.67$ at $\lambda=633\:$nm (Fig.~\ref{fig:piaa_perf_geo}), and normalizing by $C = F_\odot/F_\oplus$ (i.e. the measured contrast):
\begin{equation}
    \rhos = 0.67\: C = F_\odot \; \dfrac{0.67}{F_\oplus} 
\end{equation}
Doing so, we put aside the effective transmission of the IFU measured in Sect.~\ref{sec:ifu_perf}.

\begin{figure*}
    \centering
    \includegraphics[width=1\textwidth, trim={1cm 0 1cm 0}]{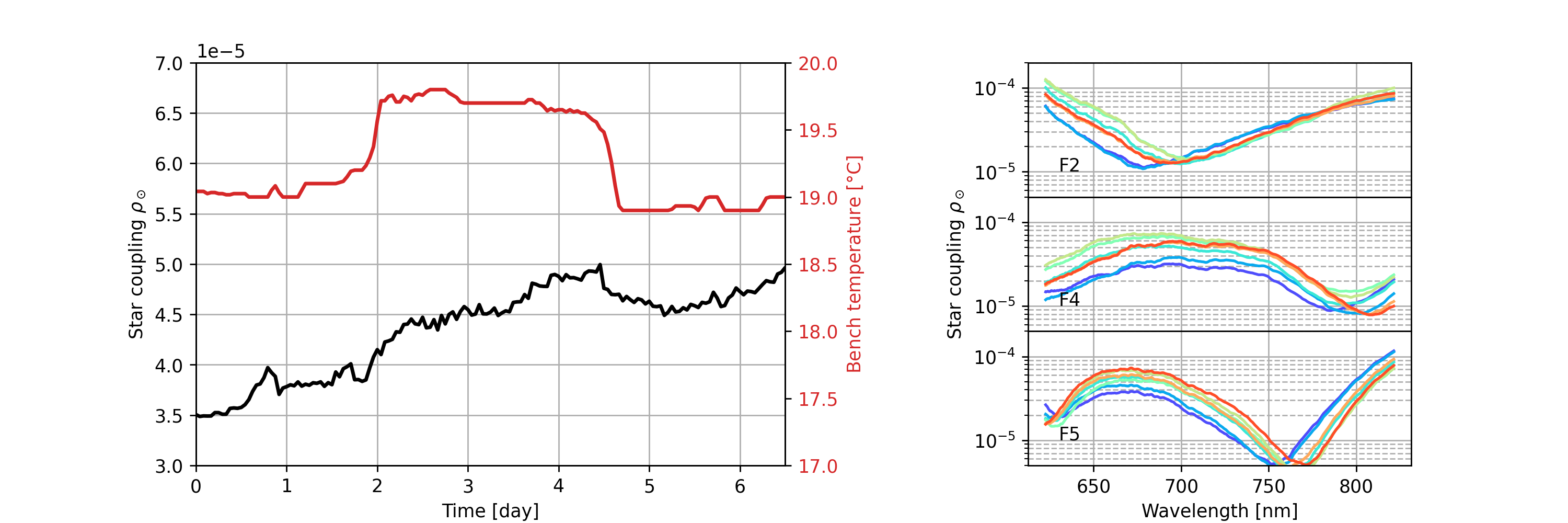}
    \caption{PIAAN stability measurement over a week, with hourly tip-tilt correction. Left: average contrast over the 6 fibers (black), and bench temperature (red). Right: contrast curves for the 3 most unstable fibers (\#2, \#4, and \#5). There is a 24h time span between each curve, from blue to red.}
    \label{fig:piaan_stability}
\end{figure*}

We start by calibrating the DM through the PIAAN to compute the eigen modes (step 1 in Sect.~\ref{sec:wfc_procedure}), and correct wavefront with the 1$^{st}$ mode. This corresponds to an additional Strehl optimization, compensating the WFE between the initial eye-doctor on the focal plane camera and the PIAAN. We could actually observe very little (if any) gain with this step, indicative of a very small amount of differential aberration between the IFU and the imaging camera. We generally obtain relatively good nulls, with $\rhosm=2.7\cdot10^{-4}$ (Fig.~\ref{fig:piaan_optim}, orange curve), which is still 5 times worse than expected from the tolerance analysis (Fig.~\ref{fig:wfc_sim}, bottom). The dispersion over wavelength and spaxels amounts to $2\cdot10^{-4}$ rms.

We then look for the working point (step 2 in Sect.~\ref{sec:wfc_procedure}). After a new calibration, we scan successively each eigen mode and apply proper amplitude to minimize $\rhos$ on each spaxel successively. The procedure is repeated several times with intermediate calibrations, converging in less than 20 iterations. With our current setup, it takes about 30\:min, limited by the reading speed of the spectrograph camera. We eventually reduce $\rhosm$ by another factor 10. With the OAP4, we obtained $\rhosm = 5.0\cdot10^{-5}$. For the reasons explained before, we exchanged this OAP for a doublet lens (Thorlabs AC254-250-A), with higher optical quality, and chromatic errors in principle below our requirements. We could then consistently obtain better contrasts, down to $\rhosm = 3.5\cdot10^{-5}$ with a dispersion of $\pm 1\cdot10^{-5}$ rms over wavelength and spaxels. We present the latter results in Fig.~\ref{fig:piaan_optim} (green curves). We attribute the lower performance of the OAP to its much higher surface roughness: since it is outside the pupil plane, the speckles created at the IFU focal plane in the 1-3\ld area can only be partially compensated by the DM.

The demonstrated performance is consistent with expectations, considering the effect of the DM structure and various errors in the system.

The full optimization procedure was repeated 80 times over 72\:h to test its robustness, starting from the same DM flat and contrast. The repeatability is excellent with a dispersion of $\pm 1\cdot10^{-5}$ rms for each spaxel and wavelength. The start point remains nearly identical from one test to the other.

The optimal pattern on the DM is shown in Fig.~\ref{fig:piaan_optim} (bottom) and is larger than anticipated in Sect.~\ref{sec:wfc}, with an optimal wavefront of 20\:nm RMS ($\sim\lambda/30-\lambda/40$), leading to a relative transmission degradation of the PIAAN of 2.6\%, or $\rhopm = 70\:\%$. Over the 80 repetitions, the DM shape varies by 1.0\:nm rms.

We also tested different pupil masks within $1\%$ of the optimal one. After the optimization, we always obtained $\rhosm \le 5\cdot10^{-5}$.

We finally performed stability tests. First a 72h long stability test has shown that contrast degradation is dominated by tip-tilt drift of the bench: after correction with the DM, the spectra of the 6 external spaxels overlapped perfectly with the initial ones. Note that the correction was applied according to the focal plane camera position, i.e. with a potentially "large" amount of differential aberration. We also performed a week long stability test (Fig.~\ref{fig:piaan_stability}, left), where only tip-tilt corrections were applied every 1\:hour, still according to the focal plane camera. Between day 2 and 5, a +1\:K temperature step was applied. The cumulated DM corrections indicate a bench drift of 0.2\:\ld, while average contrast correlates to temperature at a rate of $\Delta\rhos/\Delta T \sim 1\cdot10^{-5} K^{-1}$. Over the week, we still observe a slow drift of contrast of $\Delta\rhos \simeq +1\cdot10^{-5}$ uncorrelated to temperature. A new contrast optimization of the PIAAN allowed to recover initial performance. Therefore, stabilizing the PIAAN environment to 1K should easily keep the PIAAN operational for a full night.

\section{Conclusion}
\label{sec:ccl}

We presented the PIAAN, a high transmission ($\rhop=0.72$), high contrast ($\rhos=7\cdot10^{-7}$) and broad band ($\Delta\lambda=30\%$) coronagraphic IFU for High Dispersion Coronagraphy, designed for observing companions orbiting at about 2\ld from their host star. It is the baseline for RISTRETTO, to characterize Proxima Cen b with the VLT.

After presenting the concept, we performed a complete tolerance analysis to evaluate its manufacturability and identify the limits between the ideal design and the feasible one. It shows an excellent robustness to tip-tilt jitter, a major limitation of XAO systems in presence of vibrations or wind shake. A wavefront optimization strategy has also been developed to manage defects that are identified (or not) in the tolerance analysis.

We finally designed and tested a full prototype that reached performance in agreement with our expectations from the initial tolerance analysis, i.e. $\rhos=3.5\cdot10^{-5}$. We also investigated the use of recently developed 3D printing capability offered by 2-photon polymerization technology to produce a compact, stable, and fully integrated IFU. It demonstrates very encouraging maturity level, especially considering our design requires high volume to be printed, at the current limit of the technology. We will continue explore capabilities of this technologies, in order to reach the maximum transmission allowed by the PIAAN. A bulk solution might also be considered otherwise.

Such coronagraph development is fundamental to fully exploit spatial resolution and collecting area of large telescopes, in particular the future generation of ELTs, for which observing time will be very costly. PIAAN performance and high tolerance allow to relax constraints on XAO system for such large apertures, as well as to limit the impact of the apparent diameter of the nearest stars with bigger telescopes.

\section*{Acknowledgement}
This work has been carried out within the framework of the National Center of Competence in Research PlanetS supported by the Swiss National Science Foundation under grants 51NF40\_182901 and 51NF40\_205606. The RISTRETTO project was partially funded through SNSF FLARE programme for large infrastructures under grants 20FL21\_173604 and 20FL20\_186177. The authors acknowledge the financial support of the SNSF. The authors would like to thank Dr. Olivier Guyon for his interest in this work and the fruitful discussions. C.M. acknowledges the support from the Swiss National Science Foundation under grant 200021\_204847 “PlanetsInTime”

\bibliography{references} 

\begin{thebibliography}{32}
\expandafter\ifx\csname natexlab\endcsname\relax\def\natexlab#1{#1}\fi

\bibitem[{{Beuzit} {et~al.}(2019){Beuzit}, {Vigan}, {Mouillet}, {Dohlen}, {Gratton}, {Boccaletti}, {Sauvage}, {Schmid}, {Langlois}, {Petit}, {Baruffolo}, {Feldt}, {Milli}, {Wahhaj}, {Abe}, {Anselmi}, {Antichi}, {Barette}, {Baudrand}, {Baudoz}, {Bazzon}, {Bernardi}, {Blanchard}, {Brast}, {Bruno}, {Buey}, {Carbillet}, {Carle}, {Cascone}, {Chapron}, {Charton}, {Chauvin}, {Claudi}, {Costille}, {De Caprio}, {de Boer}, {Delboulb{\'e}}, {Desidera}, {Dominik}, {Downing}, {Dupuis}, {Fabron}, {Fantinel}, {Farisato}, {Feautrier}, {Fedrigo}, {Fusco}, {Gigan}, {Ginski}, {Girard}, {Giro}, {Gisler}, {Gluck}, {Gry}, {Henning}, {Hubin}, {Hugot}, {Incorvaia}, {Jaquet}, {Kasper}, {Lagadec}, {Lagrange}, {Le Coroller}, {Le Mignant}, {Le Ruyet}, {Lessio}, {Lizon}, {Llored}, {Lundin}, {Madec}, {Magnard}, {Marteaud}, {Martinez}, {Maurel}, {M{\'e}nard}, {Mesa}, {M{\"o}ller-Nilsson}, {Moulin}, {Moutou}, {Orign{\'e}}, {Parisot}, {Pavlov}, {Perret}, {Pragt}, {Puget}, {Rabou}, {Ramos}, {Reess}, {Rigal}, {Rochat}, {Roelfsema}, {Rousset},
  {Roux}, {Saisse}, {Salasnich}, {Santambrogio}, {Scuderi}, {Segransan}, {Sevin}, {Siebenmorgen}, {Soenke}, {Stadler}, {Suarez}, {Tiph{\`e}ne}, {Turatto}, {Udry}, {Vakili}, {Waters}, {Weber}, {Wildi}, {Zins}, \& {Zurlo}}]{beuzit_2019a}
{Beuzit}, J.~L., {Vigan}, A., {Mouillet}, D., {et~al.} 2019, \aap, 631, A155

\bibitem[{Blind {et~al.}(2024)Blind, Shinde, Dinis, Restori, Chazelas, Fusco, Guyon, K{\"u}hn, Lovis, Martinez, Motte, Sauvage, \& Spang}]{blind_2024a}
Blind, N., Shinde, M., Dinis, I., {et~al.} 2024, in Adaptive Optics Systems IX, ed. K.~J. Jackson, D.~Schmidt, \& E.~Vernet, Vol. 13097, International Society for Optics and Photonics (SPIE), 130976U

\bibitem[{Breckinridge {et~al.}(2015)Breckinridge, Lam, \& Chipman}]{breckinridge_2015a}
Breckinridge, J.~B., Lam, W. S.~T., \& Chipman, R.~A. 2015, Publications of the Astronomical Society of the Pacific, 127, 445

\bibitem[{Conan \& Correia(2014)}]{conan_2014a}
Conan, R. \& Correia, C. 2014, in Adaptive Optics Systems IV, ed. E.~Marchetti, L.~M. Close, \& J.-P. V{\'e}ran, Vol. 9148, International Society for Optics and Photonics (SPIE), 91486C

\bibitem[{Echeverri {et~al.}(2019{\natexlab{a}})Echeverri, Ruane, Jovanovic, Hayama, Delorme, Pezzato, Bond, Wang, Mawet, Wallace, \& Eugene}]{echeverri_2019b}
Echeverri, D., Ruane, G., Jovanovic, N., {et~al.} 2019{\natexlab{a}}, in Techniques and {Instrumentation} for {Detection} of {Exoplanets} {IX}, ed. S.~B. Shaklan (Proc. SPIE), 33

\bibitem[{Echeverri {et~al.}(2019{\natexlab{b}})Echeverri, Ruane, Jovanovic, Mawet, \& Levraud}]{echeverri_2019a}
Echeverri, D., Ruane, G., Jovanovic, N., Mawet, D., \& Levraud, N. 2019{\natexlab{b}}, Opt. Lett., 44, 2204

\bibitem[{Echeverri {et~al.}(2023)Echeverri, Xuan, Jovanovic, Ruane, Delorme, Mawet, Mennesson, Serabyn, Wallace, Wang, Ruffio, Finnerty, Xin, Millar-Blanchaer, Baker, Bartos, Calvin, Cetre, Doppmann, Fitzgerald, Hillman, Horstman, Hsu, Liberman, Lopez, Morris, Pezzato, Phillips, Ren, Sappey, Schofield, Skemer, Vancil, \& Wang}]{echeverri_2023a}
Echeverri, D., Xuan, J., Jovanovic, N., {et~al.} 2023, Journal of Astronomical Telescopes, Instruments, and Systems, 9, 035002

\bibitem[{Give'on {et~al.}(2007)Give'on, Kern, Shaklan, Moody, \& Pueyo}]{giveon_2007}
Give'on, A., Kern, B., Shaklan, S., Moody, D.~C., \& Pueyo, L. 2007, in Astronomical {Adaptive} {Optics} {Systems} and {Applications} {III}, Vol. 6691 (International Society for Optics and Photonics), 66910A

\bibitem[{Guyon {et~al.}(2005)Guyon, Pluzhnik, Galicher, Martinache, Ridgway, \& Woodruff}]{guyon_2005a}
Guyon, O., Pluzhnik, E.~A., Galicher, R., {et~al.} 2005, The Astrophysical Journal, 622, 744

\bibitem[{Haffert {et~al.}(2020)}]{haffert_2020a}
Haffert, S.~Y. {et~al.} 2020, Astronomy and astrophysics, 635, A56

\bibitem[{Jovanovic {et~al.}(2015)Jovanovic, Martinache, Guyon, Clergeon, Singh, Kudo, Garrel, Newman, Doughty, Lozi, Males, Minowa, Hayano, Takato, Morino, Kuhn, Serabyn, Norris, Tuthill, Schworer, Stewart, Close, Huby, Perrin, Lacour, Gauchet, Vievard, Murakami, Oshiyama, Baba, Matsuo, Nishikawa, Tamura, Lai, Marchis, Duchene, Kotani, \& Woillez}]{jovanovic_2015a}
Jovanovic, N., Martinache, F., Guyon, O., {et~al.} 2015, {\textbackslash}pasp, 127, 890

\bibitem[{{Jovanovic, N.} {et~al.}(2017){Jovanovic, N.}, {Schwab, C.}, {Guyon, O.}, {Lozi, J.}, {Cvetojevic, N.}, {Martinache, F.}, {Leon-Saval, S.}, {Norris, B.}, {Gross, S.}, {Doughty, D.}, {Currie, T.}, \& {Takato, N.}}]{jovanovic_2017a}
{Jovanovic, N.}, {Schwab, C.}, {Guyon, O.}, {et~al.} 2017, A\&A, 604, A122

\bibitem[{K{\"u}hn {et~al.}(2022)K{\"u}hn, Blind, Chazelas, Hocini, \& Lovis}]{kuhn_2022a}
K{\"u}hn, J.~G., Blind, N., Chazelas, B., Hocini, E., \& Lovis, C. 2022, in Advances in Optical and Mechanical Technologies for Telescopes and Instrumentation V, ed. R.~Navarro \& R.~Geyl, Vol. 12188, International Society for Optics and Photonics (SPIE), 121881T

\bibitem[{Liberman {et~al.}(2024)Liberman, Llop-Sayson, Bertrou-Cantou, Mawet, Desai, Haffert, \& Riggs}]{liberman_2024a}
Liberman, J., Llop-Sayson, J., Bertrou-Cantou, A., {et~al.} 2024, Journal of Astronomical Telescopes, Instruments, and Systems, 10, 029002

\bibitem[{Lovis {et~al.}(2024)Lovis, Blind, Chazelas, Shinde, Bugatti, Restori, Dinis, Genolet, Hughes, Sordet, Schnell, Rihs, Crausaz, Turbet, Billot, Fusco, Neichel, Sauvage, Diaz, Houelle, Blackman, Lanotte, K{\"u}hn, Hagelberg, Guyon, Martinez, Spang, Mordasini, Ehrenreich, Demory, \& Bolmont}]{lovis_2024a}
Lovis, C., Blind, N., Chazelas, B., {et~al.} 2024, in Ground-based and Airborne Instrumentation for Astronomy X, ed. J.~J. Bryant, K.~Motohara, \& J.~R.~D. Vernet, Vol. 13096, International Society for Optics and Photonics (SPIE), 130961I

\bibitem[{{Lovis} {et~al.}(2017)}]{lovis_2016a}
{Lovis}, C. {et~al.} 2017, Astronomy and astrophysics, 599, A16

\bibitem[{Lozi {et~al.}(2009)Lozi, Martinache, \& Guyon}]{lozi_2009a}
Lozi, J., Martinache, F., \& Guyon, O. 2009, Publications of the Astronomical Society of the Pacific, 121, 1232

\bibitem[{Macintosh {et~al.}(2014)Macintosh, Graham, Ingraham, Konopacky, Marois, Perrin, Poyneer, Bauman, Barman, Burrows, Cardwell, Chilcote, Rosa, Dillon, Doyon, Dunn, Erikson, Fitzgerald, Gavel, Goodsell, Hartung, Hibon, Kalas, Larkin, Maire, Marchis, Marley, McBride, Millar-Blanchaer, Morzinski, Norton, Oppenheimer, Palmer, Patience, Pueyo, Rantakyro, Sadakuni, Saddlemyer, Savransky, Serio, Soummer, Sivaramakrishnan, Song, Thomas, Wallace, Wiktorowicz, \& Wolff}]{macintosh_2014a}
Macintosh, B., Graham, J.~R., Ingraham, P., {et~al.} 2014, Proceedings of the National Academy of Sciences, 111, 12661

\bibitem[{Males {et~al.}(2024)Males, Close, Haffert, Kautz, Kueny, Long, McEwen, Swimmer, Iii, Foster, Mazin, Pearce, Liberman, Twitchell, Weinberger, Guyon, Hedglen, McLeod, Roberts, Gorkom, Li, Doty, \& Gasho}]{males_2024a}
Males, J.~R., Close, L.~M., Haffert, S.~Y., {et~al.} 2024, in Adaptive {Optics} {Systems} {IX}, Vol. 13097 (Proc. SPIE), 47--56

\bibitem[{Marcuse(1973)}]{marcuse_1973a}
Marcuse, D. 1973, The Bell System Technical Journal, 52, 817

\bibitem[{Mawet {et~al.}(2012)Mawet, Pueyo, Lawson, Mugnier, Traub, Boccaletti, Trauger, Gladysz, Serabyn, Milli, Belikov, Kasper, Baudoz, Macintosh, Marois, Oppenheimer, Barrett, Beuzit, Devaney, Girard, Guyon, Krist, Mennesson, Mouillet, Murakami, Poyneer, Savransky, V{\'e}rinaud, \& Wallace}]{mawet_2012a}
Mawet, D., Pueyo, L., Lawson, P., {et~al.} 2012, in Space Telescopes and Instrumentation 2012: Optical, Infrared, and Millimeter Wave, ed. M.~C. Clampin, G.~G. Fazio, H.~A. MacEwen, \& J.~M.~O. Jr., Vol. 8442, International Society for Optics and Photonics (Proc. SPIE), 844204

\bibitem[{Mawet {et~al.}(2016)Mawet, Wizinowich, Dekany, Chun, Hall, Cetre, Guyon, Wallace, Bowler, Liu, Ruane, Serabyn, Bartos, Wang, Vasisht, Fitzgerald, Skemer, Ireland, Fucik, Fortney, Crossfield, Hu, \& Benneke}]{mawet_2016a}
Mawet, D., Wizinowich, P., Dekany, R., {et~al.} 2016, in Proc. SPIE,, Vol. 9909, 99090D

\bibitem[{Mennesson {et~al.}(2006)Mennesson, Haguenauer, Serabyn, \& Liewer}]{mennesson_2006a}
Mennesson, B., Haguenauer, P., Serabyn, E., \& Liewer, K. 2006, in Advances in {Stellar} {Interferometry}, Vol. 6268, International Society for Optics and Photonics (SPIE), 944--950

\bibitem[{Por {et~al.}(2018)}]{por_2018a}
Por, E.~H. {et~al.} 2018, in Adaptive {Optics} {Systems} {VI}, Vol. 10703 (International Society for Optics and Photonics), 1070342

\bibitem[{Restori {et~al.}(2024)Restori, Blind, Kühn, Chazelas, Lovis, Mordasini, Shinde, Martinez, \& Guyon}]{restori_2024a}
Restori, N., Blind, N., Kühn, J., {et~al.} 2024, in Advances in {Optical} and {Mechanical} {Technologies} for {Telescopes} and {Instrumentation} {VI}, Vol. 13100, International Society for Optics and Photonics (SPIE), 853--863

\bibitem[{Ruane {et~al.}(2018)Ruane, Wang, Mawet, Jovanovic, Delorme, Mennesson, \& Wallace}]{ruane_2018a}
Ruane, G., Wang, J., Mawet, D., {et~al.} 2018, {\textbackslash}apj, 867, 143

\bibitem[{Ruilier(1998)}]{ruilier_1998a}
Ruilier, C. 1998, in Astronomical Interferometry, ed. {R. D. Reasenberg}, Vol. 3350 (Proc. SPIE,), 319--329

\bibitem[{{Snellen} {et~al.}(2015){Snellen}, {de Kok}, {Birkby}, {Brandl}, {Brogi}, {Keller}, {Kenworthy}, {Schwarz}, \& {Stuik}}]{snellen_2015a}
{Snellen}, I., {de Kok}, R., {Birkby}, J.~L., {et~al.} 2015, \aap, 576, A59

\bibitem[{Sparks \& Ford(2002)}]{sparks_2002a}
Sparks, W.~B. \& Ford, H.~C. 2002, The Astrophysical Journal, 578, 543

\bibitem[{{Vigan, A.} {et~al.}(2022){Vigan, A.}, {Dohlen, K.}, {N’Diaye, M.}, {Cantalloube, F.}, {Girard, J. H.}, {Milli, J.}, {Sauvage, J.-F.}, {Wahhaj, Z.}, {Zins, G.}, {Beuzit, J.-L.}, {Caillat, A.}, {Costille, A.}, {Le Merrer, J.}, {Mouillet, D.}, \& {Tourenq, S.}}]{vigan_2022a}
{Vigan, A.}, {Dohlen, K.}, {N’Diaye, M.}, {et~al.} 2022, A\&A, 660, A140

\bibitem[{Xin {et~al.}(2024)Xin, Echeverri, Jovanovic, Mawet, Leon-Saval, Amezcua-Correa, Yerolatsitis, Fitzgerald, Gatkine, Kim, Lin, Norris, Ruane, \& Sallum}]{xin_2024a}
Xin, Y., Echeverri, D., Jovanovic, N., {et~al.} 2024, Journal of Astronomical Telescopes, Instruments, and Systems, 10, 025001

\bibitem[{{Xin} {et~al.}(2022){Xin}, {Jovanovic}, {Ruane}, {Mawet}, {Fitzgerald}, {Echeverri}, {Lin}, {Leon-Saval}, {Gatkine}, {Kim}, {Norris}, \& {Sallum}}]{xin_2022a}
{Xin}, Y., {Jovanovic}, N., {Ruane}, G., {et~al.} 2022, \apj, 938, 140

\end{thebibliography}
\bibliographystyle{aa} 

\end{document}